\documentclass[prb, twocolumn, showpacs, floatfix]{revtex4-1}

\usepackage{latexsym}
\usepackage{graphicx}
\usepackage[utf8]{inputenc}
\usepackage[english]{babel}
\usepackage{amsfonts}
\usepackage{amsmath}
\usepackage{multirow}

\usepackage{amssymb}
\usepackage{dsfont}
\usepackage{braket}
\usepackage{dcolumn}
\usepackage{url}

\usepackage{color}
\usepackage{hyperref}
\hypersetup{colorlinks=true, linkcolor=blue, citecolor=black}

\newcommand{\si}{\sigma}

\newcommand{\ep}{\epsilon}

\newcommand{\up}{\uparrow}
\newcommand{\down}{\downarrow}

\newcommand{\can}[1]{\operatorname{\hat c}_{#1}}
\newcommand{\cdag}[1]{\operatorname{\hat c}^{\dagger}_{#1}}

\newcommand{\s}{\operatorname{\hat S}}
\newcommand{\sdag}{\operatorname{\hat S}^{\dagger}}

\newcommand{\gan}[1]{\operatorname{\hat \gamma}_{#1}}
\newcommand{\gdag}[1]{\operatorname{\hat \gamma}_{#1}^{\dagger}}

\newcommand{\dan}[1]{\operatorname{\hat d}_{#1}}
\newcommand{\dddag}[1]{\operatorname{\hat d}^{\dagger}_{#1}}
\newcommand{\n}[1]{\hat n_{#1}}

\newcommand{\ro}[1]{ { \hat\rho }_{#1}  }

\newcommand{\U}[2]{\operatorname{\hat U_{#1}}(#2)}
\newcommand{\Udag}[2]{\operatorname{\hat U_{#1}^{\dagger} }(#2)}

\newcommand{\trb}{\operatorname{Tr_{B} }}
\newcommand{\tr}{\operatorname{Tr}}

\newcommand{\ih}{\frac{i}{\hbar}}

\newcommand{\ketbra}[1]{~\mathinner{|{#1}\rangle\langle{#1}|}~}

\newcommand{\sgn}{\operatorname{sgn}}

\definecolor{darkgreen}{rgb}{0,0.4,0.2}

\begin{document}

\title{Subgap features due to thermally excited quasiparticles in quantum dots
coupled to superconducting leads}

\author{Sebastian Pfaller}   
\email{sebastian1.pfaller@physik.uni-r.de}
\author{Andrea Donarini}
\author{Milena Grifoni}
\affiliation{Theoretische Physik, Universität Regensburg, 93040 Regensburg,
Germany}

\date{\today}

\begin{abstract}
 We present a microscopic theory of transport through quantum dot set-ups
coupled to superconducting leads.  We derive a master equation for the reduced
density matrix to lowest order in the tunneling Hamiltonian and focus on
quasiparticle tunneling. For high enough temperatures transport occurs in the
subgap region due to thermally excited quasiparticles, which can be used to
observe  excited states of the system at low bias voltages. 
On the example of a double quantum dot we show how subgap transport spectroscopy
can be done. Moreover, we use the single level quantum dot coupled to a
normal and a superconducting lead to give a possible explanation for the subgap
features observed in the experiments of Ref. \onlinecite{Dirks2009}.
\end{abstract}

\maketitle

\section{Introduction}
\label{intro}

In the last two decades modern fabrication techniques made it possible to
connect quantum dot systems with superconducting leads. Quantum dots
were realized with  carbon nanotubes
\cite{Buitelaar2003,Vecino2004,Eichler2007,Grove-Rasmussen2009,Dirks2009,
Herrmann2010, Pillet2010}, metallic particles \cite{Ralph},
semiconducting
nanowires \cite{vanDam2006,Doh2008,Hofstetter2009,Herrmann2012arXiv}, single
fullerene molecules
\cite{Winkelmann2009}, self-assembled nanocrystals \cite{Katsaros2010} and
graphene quantum dots \cite{Dirks2011}.
The experiments show a gap in the Coulomb diamonds which is proportional
to the superconducting gap, reflecting the BCS-density of states. In the 
sequential tunneling regime higher order quasiparticle tunneling processes are
suppressed and current flows due to single quasiparticle tunneling.
First transport theories were presented
\cite{Whan1996},
using a master equation approach, where the
rates were calculated on the basis of Fermi's golden-rule. Another method
based on non-equilibrium Green's function was used by Yeyati \textit{ et al.}
\cite{Levy_Yeyati1996} and  Kang \cite{Kang1997} to describe resonant
tunneling through an effective
single level quantum dot in the limit of very strong Coulomb repulsion in the
dot ($U \to \infty$ limit), where transport is governed by quasiparticle
tunneling; the corresponding 
I-V curves show an intrinsic broadening of the BCS-like feature in the current
 in agreement with  experimental observation \cite{Ralph}.
For small Coulomb repulsion, higher order processes lead to  Josephson
current\cite{vanDam2006} and Andreev reflections
\cite{Buitelaar2003,Vecino2004,Eichler2007,Doh2008,Grove-Rasmussen2009,
Pillet2010,Dirks2011}, which appear as subgap features in the experiments.
 Both effects were studied intensely experimentally
 and theoretically
\cite{Levy_Yeyati1996,Pala2007,Eichler2007,Governale2008}
 and were recently summarized  in review articles of Refs.
\onlinecite{DeFranceschi2010,Martin-Rodero2011}.
Besides Andreev reflections also the Kondo effect \cite{Winkelmann2009} as well
as Yu-Shiba-Rusinov bound states
\cite{Grove-Rasmussen2009,Andersen2011,Franke2011} can lead to subgap features
and are the subject of current research.
If the temperature becomes comparable with the superconducting gap
quasiparticles can get thermally excited across the gap, leading to additional
subgap features \cite{Whan1996}.

 In the
following we present a microscopic theory for transport through superconducting
hybrid nanojunctions for finite superconducting gap $|\Delta| < \infty$
in the sequential tunneling limit.
In particular,  we trace out all degrees of freedom of the superconducting leads
to obtain  a generalized master equation for the
reduced density matrix to lowest order in the tunneling Hamiltonian.
{We differentiate from Ref.~\onlinecite{Whan1996}
by going beyond the constant  interaction implicitly used there, and
from
Refs.
\onlinecite{Levy_Yeyati1996} and \onlinecite{Kang1997} since we also treat
 subgap features associated to many-body excitations of a quantum dot
molecule  (double quantum dot).}
In contrast to  Green's function techniques, see e.g. Ref.
\onlinecite{Martin-Rodero2011},
this method enables one to treat the interactions on the system \textit{exactly}.
Moreover, as shown on the example of a double quantum dot,
our theory  is  easily scalable and allows an exact
treatment of the Coulomb interaction and can treat any quantum dot set-up.
Hence, we can describe lowest order quasiparticle transport of
experimental relevant quantum dot systems (multiple quantum dots or
multilevel quantum dots).
We focus on transport involving thermally excited quasiparticles, and show that
excited states of the quantum dot system can be observed in the current voltage
spectroscopy in the Coulomb blockade region. Though transitions between two
ground states are blocked due to the gap in the  BCS-density of states,
thermally excited
quasiparticles can participate in transport through excited system states,
giving
a source of subgap features in superconducting hybrid systems.
 These subgap features are already present in lowest order of
the perturbation theory,
in contrast to Cooper pair transport which occurs only in
 fourth order in the tunneling coupling. Nevertheless, experiments
suggest the existence of a regime in which quasiparticle
transport dominates also in the subgap region \cite{DeFranceschi2010}.
For a quantum dot coupled to a normal and a superconducting lead, a possible
explanation for the subgap features observed in Ref. \onlinecite{Dirks2009} is
given, where a carbon nanotube quantum dot is coupled to a normal and a
superconducting contact.

The paper is organized as follows: In Sect.~\ref{sect:model hamiltonian}
we introduce the  Hamiltonian in a system-bath model using a number conserving
version of the Bogoliubov-Valatin transformation
\cite{Josephson1962,Bardeen1962}. We describe the electrons of the
superconducting leads as a combination of  quasiparticle excitations of
the BCS-ground state and Cooper pairs. For this purpose we introduce Cooper
pair creation and annihilation operators.
The explicit inclusion of these operators allows one to construct a theory which conserves the
particle
 number in the tunneling process. In this way, for example, anomalous
 contributions to the tunneling rates due to Cooper pairing naturally
 vanish in second order.
In Sect.~\ref{sect:transport theory and the gme}, the generalized master
equation
for the reduced density matrix is derived and used to calculate the current.
In Sect.~\ref{sect:transport through multiple quantum dot devices} we apply the
theory to the calculation of transport characteristics of
two systems: the single level quantum dot (SD) and the double quantum dot (DD),
the latter in two possible configurations
cf. Fig.~\ref{fig:DD_configuration}.
The SD is used to explain basic phenomena such as a gap opening in the
Coulomb diamonds which is proportional to the superconducting gap,
and transport involving thermally excited quasiparticles \cite{Whan1996}. 
 On the other hand, the DD possess a
richer many-body spectrum with several excited states. We visualize transitions
through excited system states in the low bias regime using thermally excited
quasiparticles. 
Due to the gap in the BCS-density of states, the ground state to ground state
transition is not allowed in all cases, leading to transport through excited
system states, appearing as peaks \textit{in} the Coulomb blockade region.
The threshold for observing excited system states in the subgap region is that
the energy difference between the excited state and its ground state must be
smaller than $2|\Delta|$. We confirmed this threshold by means of the
independently gated DD, where the detuning of the two sites changes the level
spacing.
Finally the N-QD-S system is investigated, where a quantum dot is coupled to a
normal and a superconducting lead. In this case only the superconducting lead
produces thermal lines in the Coulomb blockade region, giving a possible
explanation for the subgap features in Ref. \onlinecite{Dirks2009}.

\section{Model Hamiltonian}\label{sect:model hamiltonian}

In the following we consider  quantum dot systems weakly  coupled to two
superconducting leads. 
The total Hamiltonian is written in a system-bath model:
\begin{equation}
 \hat H = \hat H_S + \hat H_B + \hat H_T,
\end{equation}
where $\hat H_S$ represents the Hamiltonian of the quantum dot system, $\hat
H_B$ is the
Hamiltonian of the superconducting leads, and $\hat H_T$ describes the tunneling
between the system and the leads.
Specifically, we focus on two systems,
a single level quantum dot (SD) and a double quantum  dot (DD). The SD
has been the focus of many theoretical works before
\cite{Whan1996,Levy_Yeyati1996,Kang1997,Pala2007,Governale2008}, and we  use its
simple Fock-space
structure to
demonstrate some generic effects resulting from the superconducting leads.

\begin{figure}
  \includegraphics[width=0.75\columnwidth]{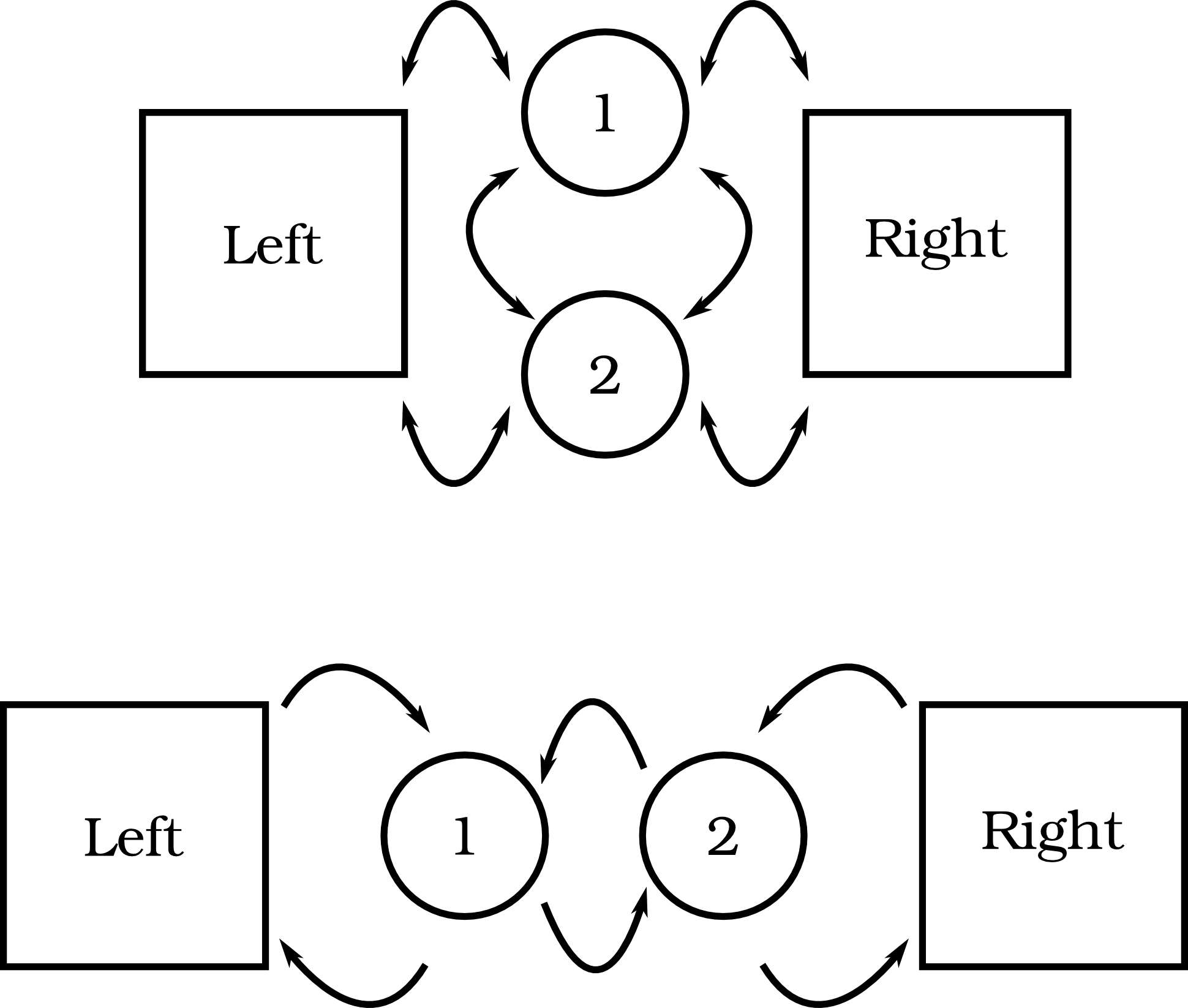}
\caption{Sketch of the transport set-up of a double quantum dot (DD) coupled to
superconducting
leads.
The DD is illustrated in the parallel (top panel) and serial (bottom panel)
configuration. Tunneling events are depicted by arrows.}
\label{fig:DD_configuration}
\end{figure}

We describe the SD by the single impurity Anderson model:
\begin{equation}\label{eq:SD}
 \hat H_{SD} = \sum_{\si} \ep_d \dddag{\si}\dan{\si} + U \n{\up}\n{\down},
\end{equation} 
where $\n{\si} = \dddag{\si}\dan{\si}$ is the number operator of the electrons
on the dot with spin $\si$. This model describes a quantum dot with on-site
energy $\ep_d$ and
Coulomb repulsion $U$ which can be occupied by at most two electrons. The
highest occupied state is defined as $\ket{2} = \dddag{\up}
\dddag{\down}\ket{0}$, the 1-particle states are defined as $\ket{1\si} =
\dddag{\si}\ket{0}$, and $\ket{0}$ is the state with zero particles.

 For the DD we use a modified version of the Pariser-Parr-Pople Hamiltonian
\cite{Pariser1953,Pople1953}:
\begin{equation}\label{eq:DD hamiltonian}
\begin{split}
&
\hat H_{DD} = \sum_{\substack{\alpha \in \{1,2 \} \\ \si \in \{ \up,\down\} }}  
\ep_{\alpha \si} \dddag{\alpha \si} \dan{\alpha \si}
+ \sum_{\si}\biggl( b \dddag{1 \si}\dan{2\si} + b^* \dddag{2\si} \dan{1\si} 
\biggr) \\
& +
\sum_{\alpha} U_{\alpha} \biggl(\n{\alpha \up} - \frac{1}{2} \biggr)
\biggl(\n{\alpha \down} - \frac{1}{2}  \biggr) + V \bigl(\n{1} -1 \bigr)(\n{2} -
1).
\end{split}
\end{equation}
Here, $\dddag{\alpha \si}$ are the creation operators for an electron  on site
$\alpha \in \{1,2\}$ with spin $\si$. They define the number operators
$\n{\alpha \si} = \dddag{\alpha \si}\dan{\alpha \si} $. The operator
$\n{\alpha} =  \n{\alpha \up } + \n{\alpha \down}$ counts the number of
electrons on site $\alpha$. In the general case we distinguish between the four
on-site energies $\ep_{\alpha \si}$ and between the on-site Coulomb
interactions $U_{\alpha}$. Electrons on different sites  interact through the
inter-dot Coulomb interaction $V$;  $b$ describes the hopping between the two
sites.
 In our set-up  the on-site energies can be controlled by  capacitively
coupled gate electrodes. In the case of site-independent on-site energies and
on-site Coulomb interaction the Hamiltonian can be diagonalized
analytically \cite{Bulka2004,Hornberger2008}.

 The superconducting leads are described by
the mean field form, $\hat H_B^{\text{MF}}$ of the pairing Hamiltonian, where we
additionally inserted
a unity represented by a product of Cooper pair  annihilation and creation
operators,
\mbox{$\s_\eta\sdag_\eta=1$}, which will be specified later in
Sec.~\ref{sect:diagonalization_lead}. We find

\begin{equation}\label{eq:MF}
\begin{split}
 &
 \hat H_B^{\text{MF}} =  \sum_{\eta k\si} \xi_{\eta k}
\cdag{\eta k \si} \can{\eta k
\si}
+\sum_{\eta} \mu_{\eta} \hat N_{\eta}\\
&	
+ \sum_{\eta k} \bigl ( \Delta_{\eta} \cdag{\eta k \up} \cdag{\eta -k \down}
\s_{\eta}
+\Delta_{\eta}^*  \sdag_{\eta} \can{\eta -k \down} \can{\eta k \up} \bigr)\\
& =
\hat H_G +\sum_\eta \mu_\eta \hat N_\eta,
\end{split}
\end{equation}
where $\xi_{\eta k} = \ep_{k} - \mu_{\eta}$ measures single particle energies
$\ep_k$ with respect to the electrochemical potential $\mu_\eta$,
 and $\hat N_{\eta} = \sum_{k \si}
\cdag{\eta k \si} \can{\eta k \si}$ counts the number of electrons in lead
$\eta$. 
 Finally,
 $\Delta_{\eta} = |\Delta_\eta|e^{i\phi_\eta} \equiv - \sum_l V_{lk}
\braket{\sdag_\eta \can{\eta -k \down} \can{\eta k \up}}$
 denotes the superconducting gap of lead $\eta$. Here $\braket{\bullet}$ denotes
a thermal average calculated self-consistently using the mean field Hamiltonian
of Eq.~(\ref{eq:MF}).

The tunneling Hamiltonian,
\begin{equation}\label{eq:tunneling Hamiltonian}
 \hat H_T = \sum_{\eta k \si \alpha} t_{\eta \alpha \si} \cdag{\eta k \si}
\dan{\alpha \si} + t^*_{\eta \alpha \si} \dddag{\alpha \si} \can{\eta k \si},
\end{equation}
describes the tunneling between the leads and the two sites of the DD, where
the tunneling coefficients $t_{\eta
\alpha \si}$ depend on the lead, site, and  spin index. Depending on the
choice of the tunneling coefficients  the DD is described in 
parallel or in serial configuration, see Fig.~\ref{fig:DD_configuration}. For
the single dot we  skip the index $\alpha$ in Eq.~(\ref{eq:tunneling
Hamiltonian}), as only one site is involved.


\subsection{Diagonalization of the lead
Hamiltonian}\label{sect:diagonalization_lead}

The most famous way to diagonalize the mean field Hamiltonian,
$\hat H_B^{\text{MF}}$, of
Eq.~(\ref{eq:MF}) was first introduced by Bogoliubov \cite{Bogoliubov1958}. We
are following Josephson and Bardeen  \cite{Josephson1962,Bardeen1962} who 
modified the so called Bogoliubov transformation in a number conserving way.
 We adopt this idea and define the Bogoliubov transformation:
\begin{equation}\label{eq:BT}
 \cdag{\eta k \si} = u_{\eta k } \gdag{\eta k \si} + \sgn{\si}\, v_{\eta k}^*
\gan{\eta -k \bar \si} \sdag_{\eta},
\end{equation}
where $\bar\si = -\si$.
In Eq.~(\ref{eq:BT}) $\gdag{\eta k \si}$ creates a fermionic quasiparticle,
often called bogoliubon, which is defined by
\begin{equation}
 \{\gdag{\eta k \si}, \gan{\eta'k'\si'}\} = \delta_{\eta \eta'} \delta_{kk'}
\delta_{\si \si'},
\end{equation}
\begin{equation}\label{eq: QPV}
 \gan{\eta k \si} \ket{\text{GS}}_{\eta} = 0.
\end{equation}
Here $\ket{\text{GS}}_{\eta} $ denotes the ground state,  or Cooper pair
condensate of lead $\eta$ \cite{note_groundstate}.
 Bogoliubons are quasiparticle excitations
of the Cooper pair condensate, meaning that the Cooper pair condensate is
defined as the vacuum state of the bogoliubons, see  Eq.~(\ref{eq: QPV}).
The coefficients $u_{\eta k}$ and $v_{\eta k }$  are complex numbers and
fulfill:
\begin{equation}\label{eq:u2+v2}
 |u_{\eta k}|^2 +|v_{\eta k}|^2 =1.
\end{equation}
They read:
\begin{equation}\label{eq:def_uk}
 u_{\eta_k} = \sqrt{ \frac{1}{2}\biggl( 1 + \frac{\xi_{\eta k}}{|E_{\eta k}|}
\biggr) },
\end{equation}
\begin{equation}\label{eq:def_vk}
 v_{\eta k} = e^{i\phi_{\eta}} \sqrt{ \frac{1}{2}\biggl( 1 - \frac{\xi_{\eta
k}}{|E_{\eta k}|} \biggr) },
\end{equation}
where $\phi_{\eta}$ is the phase of the superconducting gap
$\Delta_{\eta}$.

In the number conserving description, the Hamiltonian of
Eq.~(\ref{eq:MF}) commutes with the particle number operator. Hence, it is
required that the ground state must be an eigenstate of the  particle number
operator. We define the ground state of lead $\eta$ as
\cite{Schrieffer,Ambegaokar} $\ket{\text{GS}}_{\eta} = \ket{0,
{N}}_{\eta}$, where $\ket{0, {N}}_{\eta}$ represents a state with
$N/2$ Cooper pairs and zero quasiparticle excitations.
The Cooper pair annihilation operator
$\s_{\eta}$ annihilates a Cooper pair in lead $\eta$  and can formally be
defined as\cite{Schrieffer}:
\begin{equation}\label{eq:definition_s}
\begin{split}
 \s_{\eta} \ket{0,N}_{\eta} & = \ket{0,N-2}_\eta, \\
\s_{\eta} \ket{{k\si}, N}_{\eta} &= \ket{{k\si},N-2}_\eta, \\
\gdag{k\si} \ket{0,N}_{\eta} &= \ket{{k\si},N}_\eta.
\end{split}
\end{equation}
 Eq.~(\ref{eq:definition_s}) implies that the Cooper
pairs and the quasiparticles are decoupled:
\begin{equation}
 \big[ \sdag_{\eta} , \gdag{k\si} \big] = 0, \quad \big[\s , \gdag{k\si} \big] =
0,
\end{equation}
and  the
Cooper pair operators have the following properties, see App.~\ref{app:bcs GS
CPO}:
\begin{equation}
\s_{\eta} \sdag_{\eta}= 1, \quad  \big[\s_{\eta},\sdag_{\eta}] =
\hat{\mathcal{P}}_{0,\eta},
\end{equation}
where $\hat{\mathcal{P}}_0$ is the projector on states
with zero Cooper pairs, and
\begin{equation}
 \big[ \hat N , \sdag \big] = 2\sdag.
\end{equation} 
Note that the transformation defined in Eq.~(\ref{eq:BT})
conserves the fermionic properties of the electron operators only if  we
restrict our Hilbert space to a subspace with more than zero Cooper pairs. In
that subspace $\s$ commutes with $\sdag$ and the Bogoliubov transformation is
well defined. 

Applying the transformation of Eq.~(\ref{eq:BT}) on Eq.~(\ref{eq:MF}) we obtain
that :
\begin{equation}\label{eq:LH}
 \hat H_B - \sum_{\eta} \mu_{\eta} \hat N_{\eta} = \sum_{\eta  k \si} E_{\eta k
} \gdag{\eta k \si} \gan{\eta k \si} +
E_G + T(\hat{\mathcal{P}}_0),
\end{equation}
where $T(\hat{\mathcal{P}}_0)$ are terms 
proportional to $\hat{\mathcal{P}}_0$. They vanish after truncating the
Hilbert space and only diagonal contributions remain.
In Eq.~(\ref{eq:LH}) $E_{\eta k} = \sqrt{\xi_{\eta k}^2 + |\Delta_{\eta}|^2}$
denotes the quasiparticle energy, and $E_G$ is a constant energy off-set, often
referred to as the energy of the Cooper pair condensate. For later reference we
 note that the term $\sum_{\eta} \mu_{\eta} \hat N_{\eta}$ is not included
in the diagonalization procedure and is still written in terms of electron
operators.

\section{Transport theory and the generalized master equation}
\label{sect:transport theory and the gme}

In this section we  derive the generalized master
equation in the presence of superconducting leads. Since the  generalized master
equation approach to transport through quantum dots has become rather
standard in recent years  (see e.g. the method article by Timm \textit{et al.}
\cite{Timm2011} or the recent paper by Koller \textit{et al.} \cite{Koller2010})
we only
go into details of the derivation of the master equation when the effect of the
superconducting leads brings significant differences with respect to
the normal conducting theory. 

The expectation value $\mathcal{O} = \braket{\hat{\mathcal{O}}} =
\tr\bigl(\hat{\mathcal{O} } \hat\rho\bigr) $ of any observable associated to an
operator $\hat{\mathcal{O}}$ can be evaluated once the total density operator
$\hat\rho$ is known, cf. Eq.~(\ref{eq:current def}) below. To this extent we
start from
the Liouville equation for the density operator in the interaction picture, see
e.g. \cite{Blum1996}:
\begin{equation}\label{eq:Liouville_I}
 i \hbar \frac{\partial}{\partial t} \ro{I}(t) = \bigl[ \hat H_{T,I}(t) ,
\ro{I}(t) \bigr].
\end{equation}
Eq.~(\ref{eq:Liouville_I}) can be formally integrated and reinserted back into
itself,
\begin{equation}
\begin{split}
 i \hbar \, \dot {\hat\rho}_{I}(t) & = \bigl[\hat H_{T,I}(t), \ro{I}(0) \bigr]\\
& 
- \ih \int_0^t dt' \biggl[ \hat H_{T,I}(t), \bigl[   \hat H_{T,I}(t'),
\ro{I}(t') \bigr]    \biggr],
\end{split}
\end{equation}
which is still exact and allows a perturbative treatment
in the
tunneling Hamiltonian $\hat H_T$.

Prior to time $t= 0$ the bath and the system do not interact, meaning that the
total density matrix is factorized into a system and a leads component:
\begin{equation}
 \ro{I}(0) = \ro{S}(0) \ro{B}(0).
\end{equation}
The density matrix of the leads, $\ro{B}$, can be described by the equilibrium
thermodynamic expression shown in Eq.~(\ref{eq:rho_B}).
Further we assume that the leads have so many degrees of freedom that
they  stay in thermal equilibrium up to a correction of order $\hat H_T$.
It is convenient to trace out the degrees of freedom of the leads and
define the reduced density matrix:
\begin{equation}
 \ro{red,I}(t) \equiv \trb \ro{I}(t).
\end{equation} 
In the Schrödinger picture, the master equation for the reduced density matrix
reads:
\begin{equation}\label{eq:red}
\begin{split}
&
\dot{\hat \rho}_{red}(t) = \ih \bigl[ \ro{red}(t), \hat H_S \bigr]
- \biggl( \ih \biggr)^2 \U{0}{t} \int_{0}^{t}dt'  \times \\
&\times
\trb \biggl(  
\biggl[ \hat H_{T,I}(t), \biggl[\hat H_{T,I}(t'), \ro{red,I}(t') \ro{B}  \biggr] 
\biggr]
\biggr) \Udag{0}{t} ,
\end{split}
\end{equation}
where we neglect terms of order $\mathcal{O}(\hat H_T^3)$ and 
$\U{0}{t}= e^{-\ih \hat H_S t}$
is the time evolution operator of the unperturbed system.

\subsection{Superconducting leads}

The features of the superconducting leads are revealed when using the
Bogoliubov transformation (\ref{eq:BT}) to express the tunneling Hamiltonian.
This yields  additional terms compared to the normal conducting theory.

\subsubsection{Thermodynamic properties of the leads}
\label{sect:thermodynamic properties of the leads}

The description of electrons in terms of bogoliubons and Cooper pairs makes it
necessary to discuss the thermodynamic properties of the superconducting leads.
In this section we drop for simplicity the lead index $\eta$, and consider only
one lead.

In order to calculate thermal expectation values we use the equilibrium density
matrix of a superconductor:
\begin{equation}\label{eq:rho_B}
 \ro{B} = \frac{e^{-\beta \hat H_G}}{Z_G},
\end{equation}
where $\hat H_G = \hat H_B -\mu \hat N$, $\beta$ is the inverse thermal
energy, and $Z_G$ is the partition function in the grand canonical ensemble.
We find that the thermal expectation value of  a pair of Bogoliubov
quasiparticles is equal to the Fermi function:
\begin{equation}\label{eq:fermi}
 \trb \biggl( \gdag{k\si}\gan{k\si}  \ro{B} \biggr) = 
\frac{1}{e^{ \beta E_k} + 1}
 =f^+(E_k),
\end{equation}
where the trace is over the many-body states
\begin{equation}\label{eq:many-body excitation state}
 \ket{ \{n_{q\tau}\},N }
=  \prod_{q\tau}( \gdag{q\tau})^{n_{q\tau}}\ket{0,N},
\end{equation} with independent sums over the number of electrons $N$ in the 
Cooper pair condensate  and the  quasiparticle configuration $\{n_{q\tau}\}=\{
n_{q_1 \tau_1}, n_{q_2 \tau_2}, \, \dots \}$.

\subsubsection{Time evolution of the quasiparticles}

To proceed we have to specify the time evolution of the Bogoliubov and
Cooper pair operators. We find:
\begin{equation}\label{eq:time_gdag}
 \gdag{\eta k \si,I}(t) = e^{+\ih (E_k + \mu_{\eta})t} \gdag{\eta k \si},
\end{equation} 
\begin{equation}\label{eq:time_sdag}
 \sdag_{\eta,I}(t) = e^{+\ih 2 \mu_{\eta}t} \sdag_{\eta},
\end{equation}
 in  agreement
with the results of Josephson and Bardeen \cite{Josephson1962,Bardeen1962}. 
When calculating the time evolution it is important to remember that in the lead
Hamiltonian the term $\mu_{\eta }\hat N_{\eta}$ is still written in
terms of electron operators.

Before we proceed, we  like to emphasize the importance of the Cooper pair
contribution for finite bias voltages.  As already pointed out by Governale
\textit{et al.} \cite{Governale2008},
in this case  $\mu_\eta$ cannot be set to zero and the time evolution of the
Cooper pair operators, Eq.~(\ref{eq:time_sdag}), plays an important role.
Neglecting the Cooper pair contribution for finite bias voltages 
\cite{Kosov2012} violates the number conservation in the tunneling processes
and can lead to coherences which would vanish in the number conserving case.

\subsubsection{Difference to the normal conducting theory}

To compute Eq.~(\ref{eq:red}) we rewrite the electron operators using
the Bogoliubov transformation, Eq.~(\ref{eq:BT}), and insert the time evolution
as in Eqs.~(\ref{eq:time_gdag}) and (\ref{eq:time_sdag}). This yields
four different traces to be calculated. We find:
\begin{equation}\label{eq:trace cdag c}
\begin{split}
&
\trb\biggl( \cdag{\eta k \si ,I }(t) \can{\eta' k' \si' ,I }(t') \ro{B} \biggr)
=
\\&
\delta_{\eta \eta'}\delta_{k k'}\delta_{\si \si'} 
 \biggl\{
|u_{\eta k}|^2 f^+(E_{\eta k}) e^{+\ih( E_{\eta k } + \mu_{\eta})(t-t')}  \\
&
\phantom{\delta_{\eta \eta'}\delta_{k k'}\delta_{\si \si'}  } +
|v_{\eta k}|^2 f^-(E_{\eta k}) e^{-\ih(E_{\eta k } - \mu_{\eta}) (t-t')}
\biggr\},
\end{split}
\end{equation}
\begin{equation}\label{eq:trace can cdag}
\begin{split}
&
\trb\biggl( \can{\eta k \si ,I }(t) \cdag{\eta' k' \si' ,I }(t') \ro{B} \biggr)
=
\\&
\delta_{\eta \eta'}\delta_{k k'}\delta_{\si \si'} 
 \biggl\{
|u_{\eta k}|^2 f^-(E_{\eta k}) e^{-\ih( E_{\eta k } + \mu_{\eta})(t-t')}  \\
&
\phantom{\delta_{\eta \eta'}\delta_{k k'}\delta_{\si \si'}  } +
|v_{\eta k}|^2 f^+(E_{\eta k}) e^{+\ih(E_{\eta k } - \mu_{\eta}) (t-t')}
\biggr\},
\end{split}
\end{equation}
\begin{equation}\label{eq:trace cdag cdag}
\begin{split}
&
 \trb\biggl( \cdag{\eta k \si ,I }(t) \cdag{\eta' k' \si' ,I }(t') \ro{B}
\biggr) = 0,
\end{split}
\end{equation}
\begin{equation}\label{eq:trace can can}
\begin{split}
&
 \trb\biggl( \can{\eta k \si ,I }(t) \can{\eta' k' \si' ,I }(t') \ro{B} \biggr)
= 0,
\end{split}
\end{equation}
where $f^-(E) = 1-f^+(E)$.
Note that the trace in Eqs.~(\ref{eq:trace cdag cdag}) and (\ref{eq:trace can
can}) are vanishing since the lead Hamiltonian, Eq.~(\ref{eq:LH}), conserves the
particle number.

\subsection{General Master Equation for the reduced density matrix}

Collecting all the previous results and expressing Eq.~(\ref{eq:red}) in the
basis of the system eigenstates, $\{ \ket{n} \}$, we obtain
the Bloch-Redfield form of the general master equation (GME) for the reduced
density matrix:
\begin{equation}\label{eq:GME}
\begin{split}
 \dot{\rho}_{nn'} &=   -\ih \bigl(E_n- E_{n'}\bigr) \rho_{nn'}(t)  \\
&- \sum_{m m'}\biggl( 
R_{nn'mm'}^{N\rightarrow N+1} + R_{nn'mm'}^{N\rightarrow N-1}    
\biggr)
\rho_{mm'}(t),
\end{split}
\end{equation}
where $n$ is a collective quantum number of the many body
states of the quantum dot system and $\rho_{nn'} \equiv \bra{n}\hat \rho_{red}
\ket{n'} $. Here, the Redfield-tensors are defined as:
\begin{equation}\label{eq:redfield tensor}
\begin{split}&
 R_{nn'mm'}^{N\rightarrow N \pm 1} =  \sum_{\eta} \biggl\{ \\&
\delta_{m'n'} \sum_{ l  } 
\bigl(\Gamma^+_{nllm} \bigr)^{N\rightarrow N\pm1}_{\eta}
+\delta_{m n} \sum_{l}
\bigl( \Gamma^-_{m'lln'} \bigr)^{N\rightarrow N\pm1}_{\eta}
\\
&  
- \bigl(\Gamma^+_{m'n'nm}\bigr)^{N\rightarrow N\pm1}_{\eta}
 - \bigl( \Gamma^-_{m'n'nm}\bigr)^{N\rightarrow N\pm1}_{\eta}
\biggr\}.
\end{split}
\end{equation}
The rates $\Gamma$ in Eq.~(\ref{eq:redfield tensor}) originate from terms
 containing traces of the type of Eqs.~(\ref{eq:trace cdag c}) and
(\ref{eq:trace can cdag}). Further, we distinguish between rates  describing
the increase and rates describing the decrease of the particle number on the
system, emphasized with the superscript $N\rightarrow N\pm 1$.
Their detailed  form is presented in App.~\ref{app:rates}.
The  rates with the superscripts $\pm$ are connected by
complex
conjugation and  reversing of the indices:
\begin{equation}\label{eq:rel gamma+-}
 \bigl( \Gamma^{-}_{n m m'n'}\bigr)_{\eta}^{N\rightarrow N\pm1} 
= 
\biggl(
 \bigl( \Gamma^{+}_{n'm' m n}\bigr)_{\eta}^{N\rightarrow N\pm1}
\biggr)^* .
\end{equation}

\subsection{Current}
Having derived the GME for the reduced density matrix in Eq.~(\ref{eq:GME}), we
can use it to calculate measurable quantities such as the
current and the differential conductance.
In this section we present an expression for the current derived
from the second order GME  of Eq.~(\ref{eq:GME}). To do this we introduce a
current operator whose statistical average gives the total current:
\begin{equation}\label{eq:current def}
 I_{\eta} = \tr \bigl( \hat I_\eta\, \ro{tot} \bigr).
\end{equation}
In general, the current operator of lead $\eta$ is defined as the variation of
the total particle number in lead $\eta$ with time:
\begin{equation}\label{eq:current operator}
\hat I_{\eta,I}(t) = - e \frac{d}{dt} \hat N_{\eta,I}(t) = \frac{ + i e}{\hbar}
\biggl[ \hat N_{\eta,I}(t) , \hat H_{T,I}(t) \biggr].
\end{equation} 
Calculating the commutator of Eq.~(\ref{eq:current operator}), we see that the
current operator has the same operatorial structure  as the
tunneling Hamiltonian:
\begin{equation}
\begin{split}
 \hat I_{\eta,I}(t) =\frac{ + i e}{\hbar} \sum_{ k \alpha } \biggl(   
t_{\eta \alpha \si} \cdag{\eta k \si,I}(t) \dan{\alpha \si,I}(t)
\\
 - t^*_{\eta
\alpha \si} \dddag{\alpha \si,I}(t) \can{\eta k \si,I}(t)
\biggr),
\end{split}
\end{equation} 
differing only in the prefactor and summation. Hence, by applying the
same perturbation theory as before, we obtain
for the current in lead $\eta$:
\begin{equation}\label{eq:current}
 I_\eta(t) = e \sum_{n m l} \biggl( 
\bigl(\Gamma_{nllm}^{N\rightarrow N+1}\bigr) _\eta - 
\bigl(\Gamma_{nllm}^{N\rightarrow N-1}\bigr)_\eta
\biggr) \, \rho_{mn}^N(t) .
\end{equation} 
In Eq.~(\ref{eq:current}) we introduced the abbreviations
\begin{equation}
\begin{split}
 \bigl(\Gamma_{nm m'n'}^{N\rightarrow N\pm1}\bigr) _\eta &= 
 \bigl(\Gamma_{nmm'n'}^+\bigr)^{N\rightarrow N\pm1}_\eta
 +\bigl(\Gamma_{nmm'n'}^-\bigr)^{N\rightarrow N\pm1}_\eta \\
&=
2\, \text{Re} \biggl(  \bigl(\Gamma_{nmm'n'}^+\bigr)^{N\rightarrow N+1}_\eta 
\biggr),
\end{split}
\end{equation} 
 exploiting Eq.~(\ref{eq:rel gamma+-}). This gives us rates which are real and
read:
\begin{equation}\label{eq:rate_N_N+1}
\begin{split}
  \bigl(\Gamma_{nm m'n'}^{N\rightarrow N + 1}\bigr) _\eta & =
\text{Re}\bigg(
\tilde\Gamma_{nmm'n'}^\eta 
~
D\big(E_{m'n'} - \mu_\eta  + i \gamma \big) \\
& \phantom{= Re} \times
 f^+\big(E_{m'n'} - \mu_\eta +i\gamma \big) \bigg)
,
\end{split}
\end{equation}
\begin{equation}\label{eq:rate_N_N-1}
\begin{split}
  \bigl(\Gamma_{nm m'n'}^{N\rightarrow N - 1}\bigr) _\eta  &=
\text{Re} \bigg(
\tilde \Gamma_{m'n'n m}^\eta 
~
D\big(E_{n'm'} - \mu_\eta + i\gamma \big)~ \\
& \phantom{= Re} \times
f^-\big(E_{n'm'} - \mu_\eta+ i\gamma \big)
\bigg),
\end{split}
\end{equation}
where
\begin{equation}
 \tilde\Gamma_{nmm'n'}^\eta=
\frac{2\pi}{\hbar} \sum_{\si \alpha \alpha'} 
t_{\eta \alpha \si} t^*_{\eta \alpha' \si}
\bra{n} \dan{\alpha \si} \ket{m}\bra{m'} \dddag{\alpha' \si} \ket{n'} .
\end{equation}
In Eqs.~(\ref{eq:rate_N_N+1}) and (\ref{eq:rate_N_N-1})  $E_{n'm'}= E_n'-E_m'$
 denote differences between system eigenenergies and
\begin{equation}\label{eq:dos}
 D(E)= \rho_N \text{Re} \bigg(\frac{|E|}{\sqrt{E^2-|\Delta|^2}} \bigg),
\end{equation}
is the BCS-density of states, with 
  $\rho_N = \frac{V m k_F}{2 \pi^2 \hbar^2}$
labeling the density of states for normal leads which
is assumed to be constant around the Fermi level; $V$ denotes the volume of the
lead and $m$ is the  electron mass.
In order to renormalize the divergence of the density of states we introduced a
finite lifetime $\hbar/\gamma$ of the quasiparticle states in the
superconducting leads,
leading to a Lorentzian broadening of the resonance condition, see
App.~\ref{app:renormalization}.
 This assumption is also in agreement with the results of Levy Yeyati
\textit{et al.} \cite{Levy_Yeyati1996}, where they showed that the broadening of
the BCS-like features in the current is due to the coupling to the leads.
Eq.~(\ref{eq:current}) is a  general result and can be applied to any transport
set-up where an arbitrary system with discrete levels is weakly coupled to
superconducting or normal conducting leads. The normal conducting case is
obtained by setting $|\Delta_{\eta}|= 0$ and $\gamma = 0$.

The theory is valid in  the so called weak coupling limit, which
is defined by the following relations between fundamental energy
scales of the system: $\Gamma \ll |\Delta| \ll U$ and $\Gamma \ll k_B T$,
where $\Gamma$ is the level broadening due to hybridization with the leads, $U$
is the charging energy, and $|\Delta|$ is the superconducting gap. As
proven for example in Ref. \onlinecite{Levy_Yeyati1996}, the inclusion of higher
order terms only produces in this regime an effective broadening of
the quasiparticle density of states without invalidating the
sequential tunneling description. 

 In this paper we are only interested in the stationary limit. Hence, we
replace the density matrix in Eq.~(\ref{eq:current}) by its stationary solution
which is
determined from Eq.~(\ref{eq:GME}) by imposing  $\dot
\rho_{nn'}^N = 0$.

\section{Transport through multiple quantum dot devices}
\label{sect:transport through multiple quantum dot devices}

In the preceding sections we developed a perturbative microscopic theory for the
stationary
current of quantum dot devices coupled to superconducting leads. In
the following, we show the predictions of the theory for two models, the
single level quantum dot (SD) and the double quantum  dot (DD).
 In the transport set-up the bias and gate voltages influence the energy
configuration of the leads and the system, respectively. Specifically, the bias
voltage is modifying the electrochemical potential of the leads, which we
choose to have a
symmetric voltage drop. Therefore we  define the chemical potentials of the
left and right lead, respectively:
\begin{equation}
 \mu_{L/R} = \mu_0 \pm e \frac{V_b}{2},
\end{equation} 
where $\mu_0$ is the equilibrium chemical potential. The gate voltages are
modifying the on-site energies of the system: We replace $\ep_d \rightarrow
\ep_d + e V_g$ in  the SD- and $\ep_{\alpha} \rightarrow \ep_{\alpha} +e
V_g^\alpha$ in
the DD-Hamiltonian. Here $e = - |e|$ is the electron charge.

In the following we neglect coherences in the GME, considering only diagonal
contributions of the reduced density matrix $\rho_{nn}$ by setting $n=n'$ in
Eq.~(\ref{eq:GME}). Hence, it suffices to use only two indices for the
transition rates.

Neglecting the coherences is a non
trivial step in the derivation of the master equation for the
system. Within the secular approximation, see Ref. \onlinecite{Blum1996},
justified in the weak coupling limit, only
coherences between degenerate states can play a role. We can now
distinguish three types of degeneracies in the many-body spectrum
of a quantum dot molecule: spin degeneracy, orbital degeneracy, and
degeneracy between states with different particle number. Spin
degeneracies can be neglected in the presence of unpolarized or
collinearly polarized leads
\cite{Hornberger2008,Donarini2010}. Orbital degeneracies are system dependent
and they are
not present in the single and double quantum dot systems discussed
in this paper. A detailed discussion of their effects can be
found for example in Refs. \onlinecite{Schultz2009,Donarini2010}. 
A detailed analysis of Eq.~(\ref{eq:GME}) shows that only 'anomalous' terms
originating from Eqs.~(\ref{eq:trace cdag cdag}) and (\ref{eq:trace can can})
could couple populations
($\rho_{N,N}$) with coherences ($\rho_{N-1,N+1}$). Since these terms
are exactly vanishing in the number conserving description of the
superconducting leads, coherences decouple from populations and
vanish in the stationary limit due to the damping introduced by
the "R" components. 
\begin{figure}
  \includegraphics[width = \columnwidth]{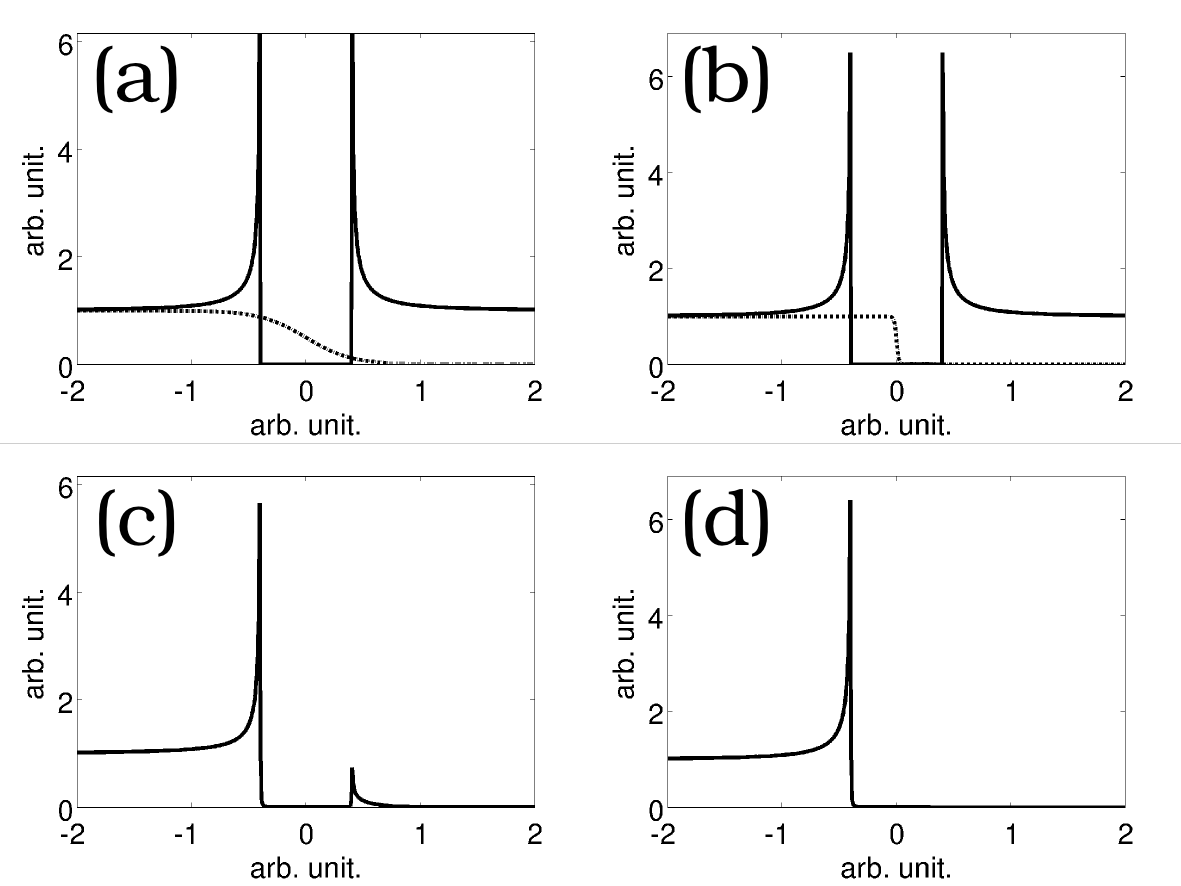}
\caption{Panels (a) and (b): Density of states (continuous line) and Fermi
function (dotted line) at $k_B T =
0.2\,$meV and $k_B T= 0.01\, $meV, respectively. Panels (c) and (d): Product of
the density of states and the Fermi function for the temperatures used in
Fig.~(a) and (b), respectively. }
\label{fig:fermi_dos}
\end{figure}
\begin{figure}
  \includegraphics[width=\columnwidth]{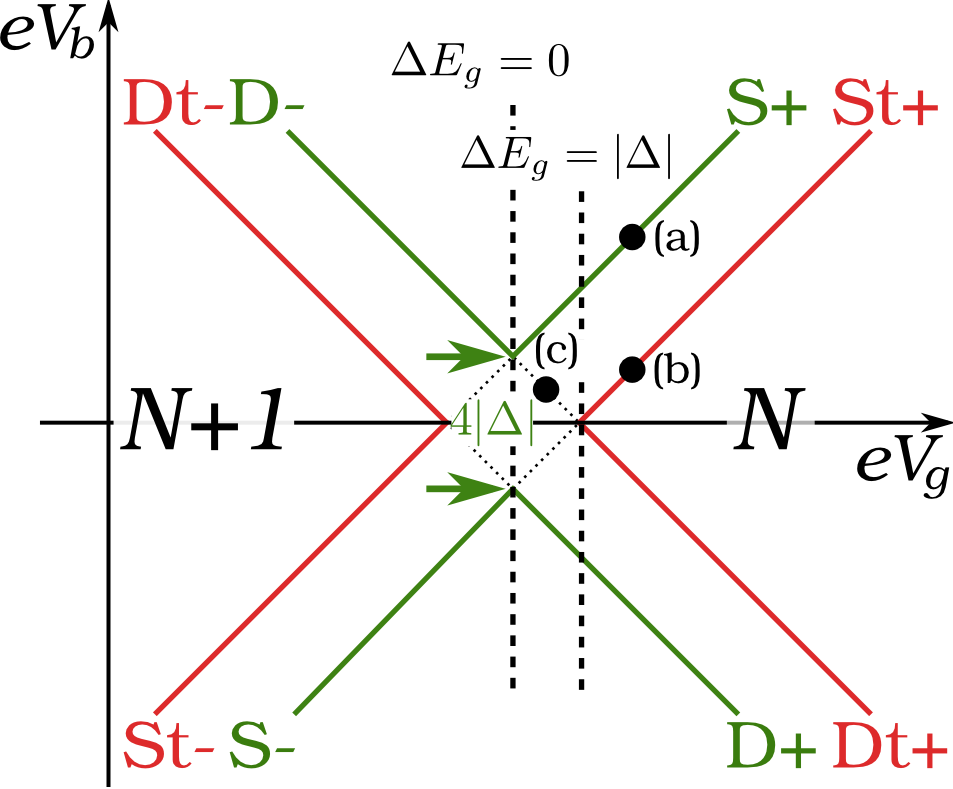}
\caption{(Color online) Illustration of the transition lines appearing in
presence of
superconducting leads. The green lines mark transitions at the \textbf Source
and the \textbf Drain contacts, described by the inequalities of
 Eqs.~(\ref{eq:s+}), (\ref{eq:d+}), (\ref{eq:s-}), and (\ref{eq:d-}). The red
lines mark transitions involving \textbf
thermally excited quasiparticles, given by Eqs.~(\ref{eq:st+}), (\ref{eq:dt+}),
(\ref{eq:st-}), and (\ref{eq:dt-}). The $E_g$-$N$ diagrams for the points
(a)-(c) are
sketched in Fig.~\ref{fig:single_level_E-N}. }
\label{fig:trans_lines}
\end{figure}

In current voltage spectroscopy it is convenient to illustrate the conditions
under which current is  allowed to flow as lines in the stability
diagrams. These so called transition lines are fixed by the energetic part
of the transition rates at the source $\eta=S$ and the drain $\eta = D$
contact:
\begin{equation}\label{eq:gamma_n_n+1}
\big( \Gamma^{N\rightarrow N + 1}_{mn}\big)_\eta \propto f^+(\Delta E -
\mu_\eta)D(\Delta E -
\mu_\eta),
\end{equation} 
\begin{equation}\label{eq:gamma_n+1_n}
\big( \Gamma^{N +1 \rightarrow N}_{nm}\big)_\eta  \propto f^-(\Delta E -
\mu_\eta)D(\Delta E
- \mu_\eta),
\end{equation} 
neglecting the lifetime broadening $\gamma$ for simplicity, and with $\Delta E =
E^{N+1}_m -E^N_n$  the energy difference of the two transport levels.
Fig.~\ref{fig:fermi_dos} illustrates
this product for two different temperatures: For high
enough temperatures quasiparticles can be excited thermally across the gap
giving a small peak in the transition rates \cite{Whan1996}.
The peak positions define transition lines when plotted in a $V_g$-$V_b$
diagram. Notice that while the most pronounced peak survives also at zero
temperature and defines a transport threshold, the second peak vanishes at low
temperatures and essentially only processes at and close to the peak are
relevant.
 For an $N\rightarrow N+1$ transition we denote 
transitions associated to the more pronounced peak as S+ and D+ when happening 
at the source or at the drain contact, respectively.
 Transitions involving thermally excited quasiparticles   are
called St+ and Dt+.
In complete analogy, we classify transitions from $N+1 \rightarrow N$:  We
denote by  S- and D- the more pronounced transitions
 at the source and at the drain, and by St- and Dt-  their thermal
counterparts.
In total we find 8 different transition lines, as depicted in
Fig.~\ref{fig:trans_lines}. In the following we derive transport conditions and
provide equations for the transport lines. For convenience we introduce
$\Delta E_g = \Delta E - \mu_0$.

We start with the analysis of the $N\rightarrow N+1$ transitions, which are
described by the rates in Eq.~(\ref{eq:gamma_n_n+1}). From the arguments we find
that the rates do not vanish
if
\begin{align}\label{eq:s+}
  {\Delta E_g \leq - |\Delta| + \frac{e V_b}{2}}, & & \text{Source {S+}}
\end{align}
\begin{align}\label{eq:d+}
 { \Delta E_g \leq - |\Delta| - \frac{e V_b}{2}}. & & \text{Drain D+}
\end{align}
Another contribution comes from the thermally excited quasiparticles states,
namely, if the argument of the Fermi function $f^+ (\Delta E - \mu_\eta)$ and of
the density of states $D(\Delta E - \mu_\eta)$ is equal to $|\Delta|$. At this
point the transition rates are peaked and  contribute to the  current:
\begin{align}\label{eq:st+}
  {\Delta E_g =  |\Delta| + \frac{e V_b}{2}}, & & \text{Source thermal St+}
\end{align}
\begin{align}\label{eq:dt+}
 { \Delta E_g =  |\Delta| - \frac{e V_b}{2}}. & & \text{Drain thermal {Dt+}}
\end{align}
Since the thermally excited quasiparticles produce a peak rather than a step in
the current voltage characteristic, the corresponding transport condition is
formulated with an equality.

Transitions from $N+1 \rightarrow N$ are described by the rate of
Eq.~(\ref{eq:gamma_n+1_n}), leading in complete analogy to the previous case to
the
following transport conditions:
\begin{align}\label{eq:s-}
 {- \Delta E_g \leq -|\Delta| - \frac{e V_b}{2}}, & & \text{Source {S-}}
\end{align}
\begin{align}\label{eq:d-}
 {- \Delta E_g \leq -|\Delta| + \frac{e V_b}{2}}, & &\text{Drain {D-}} 
\end{align}
\begin{align}\label{eq:st-}
 {- \Delta E_g = |\Delta| - \frac{e V_b}{2}}, & & \text{Source thermal {St-}} 
\end{align}
\begin{align}\label{eq:dt-}
 {- \Delta E_g = |\Delta| + \frac{e V_b}{2}}. & & \text{Drain thermal
{Dt-}}
\end{align}

\subsubsection{Visualization of the transport conditions}
\begin{figure}
\includegraphics[width = \columnwidth]{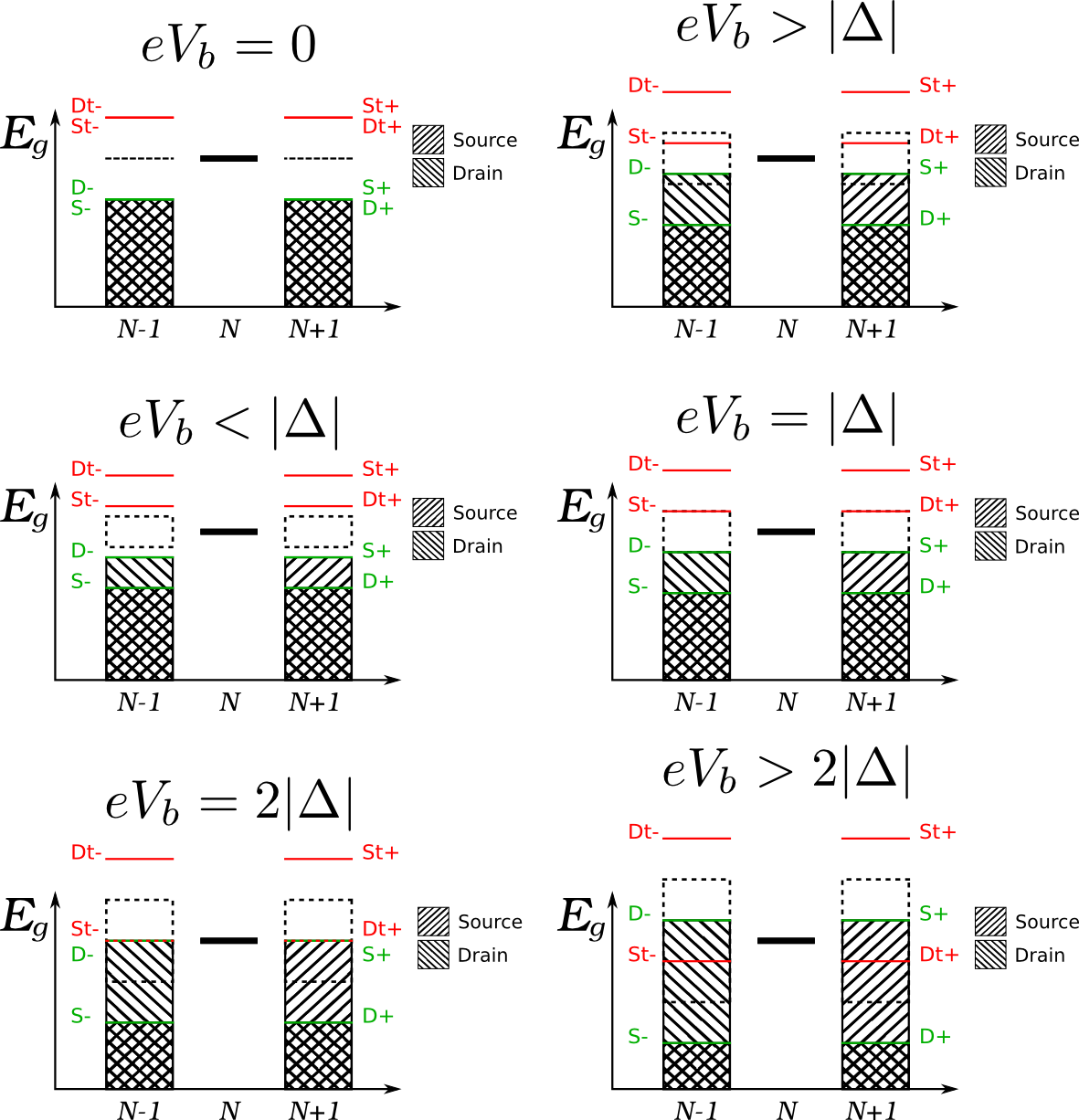}
\caption{(Color online) Visualization of the transport conditions of
Eqs.~(\ref{eq:s+})-(\ref{eq:dt-}). We plotted the threshold of the transport
inequalities as green lines (S$\pm$, D$\pm$);  for the equalities coming from
transitions involving thermally excited quasiparticles we used red lines
(St$\pm$, Dt$\pm$ ). Choosing the reference level in the $N$
particle subspace, we found a scheme where transitions are energetically allowed
to levels which lie in the shaded region below the green lines and to levels
which align with the red lines. Dashed boxes mark the bias window
$eV_b$. }
\label{fig:scheme}
\end{figure}

To visualize the transport conditions of Eqs.~(\ref{eq:s+})-(\ref{eq:dt-}) we
extend the scheme of Donarini \textit{et al.} of Ref. \onlinecite{Donarini2010}
to superconducting leads.
 The scheme is depicted in Fig.~\ref{fig:scheme} and
illustrates for which relative position of the systems eigenenergies 
$E_g^N = E_m^N - \mu_0 N$ transitions
are energetically allowed.  The bias window is marked with a
dashed box.
 The green lines mark the borders of the 
inequalities, and the red lines the sharp equalities for the thermal transitions,
meaning that transitions can occur to  states lying below the green lines
(shaded region),
and to states which coincide with the red lines. In order to see a transition
between two levels in the stability diagram a source and a drain transition
must be allowed between the two levels (depicted as arrows in the $E_g$-$N$
diagrams of Fig.~\ref{fig:single_level_E-N}).
We  note that for a full analysis of the transport properties also the
geometrical part of the rates must be taken into account and transport occurs
only if $\tilde \Gamma \neq 0$.

\subsection{ Single level quantum dot  model}

\begin{figure}
\includegraphics[width = \columnwidth]{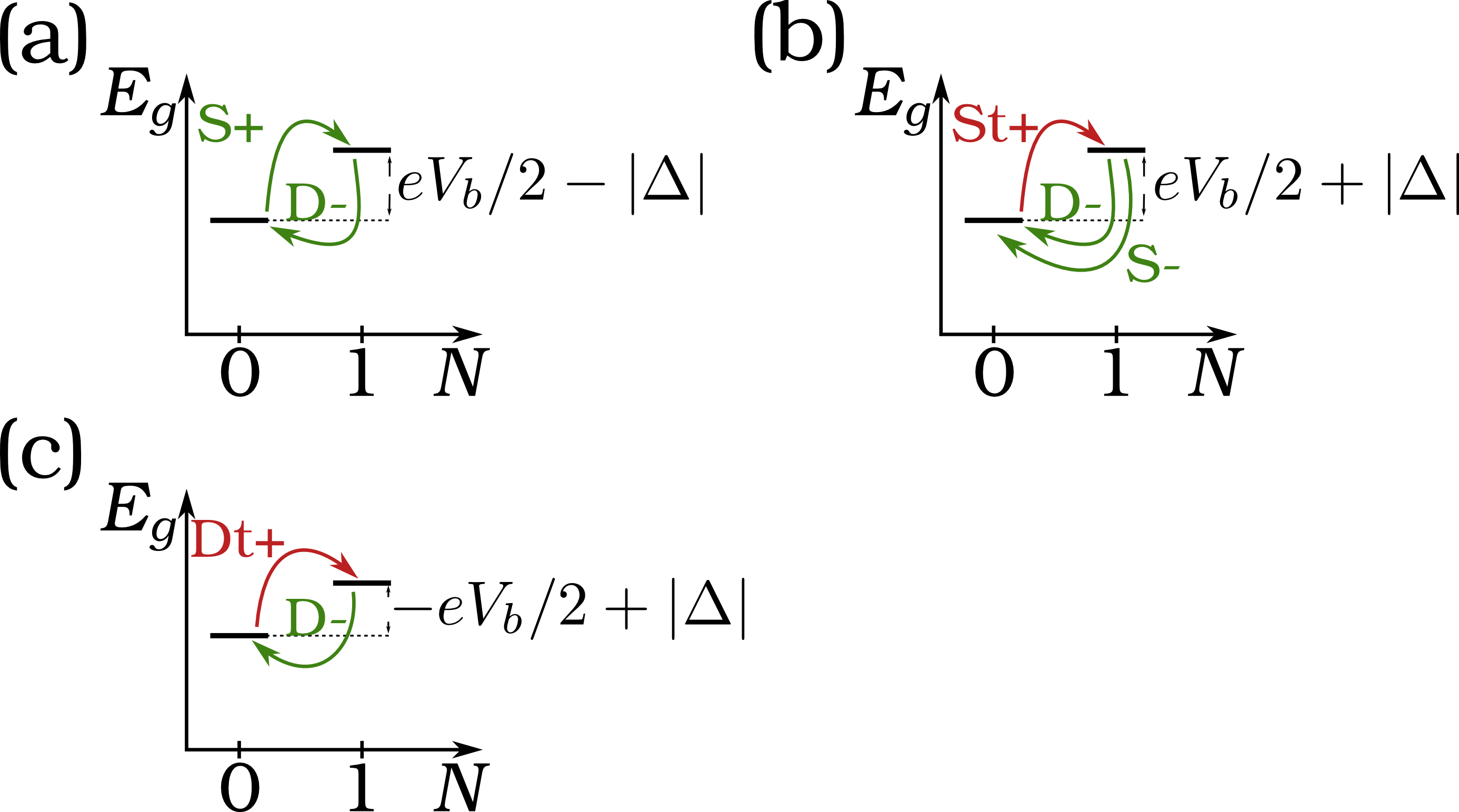}
\caption{(Color online) \textbf{(a)-(d)}: $E_g$-$N$
diagrams for a single level
quantum dot with $\Delta E_g > |\Delta|$ 
and at bias voltages as sketched in Fig.~\ref{fig:trans_lines}. For the
simulations of Fig.~\ref{fig:SD_LT} $\Delta E_g > \Delta$
corresponds to a gate voltage  $eV_g < -2.6 \,\text{meV}$. In (a) we cut the S+
line: the particle number on the system is
increased by a tunneling event at the source contact and decreased at the drain.
(b) Cut with the thermal line St+: the particle number of the system is
increased by a tunneling event involving a thermally excited quasiparticle at
the source contact and decreased by tunneling into empty states in the source
and the drain contact, respectively. 
\textbf{(c)}: $E_g$-$N$ diagrams for a single level with $0<\Delta E_g <
|\Delta|$.
The two levels are only connected by two drain transitions, meaning that in this
configuration the system is in thermal equilibrium with the drain contact.
}
\label{fig:single_level_E-N}
\end{figure}

\begin{figure}
  \includegraphics[width =  \columnwidth]{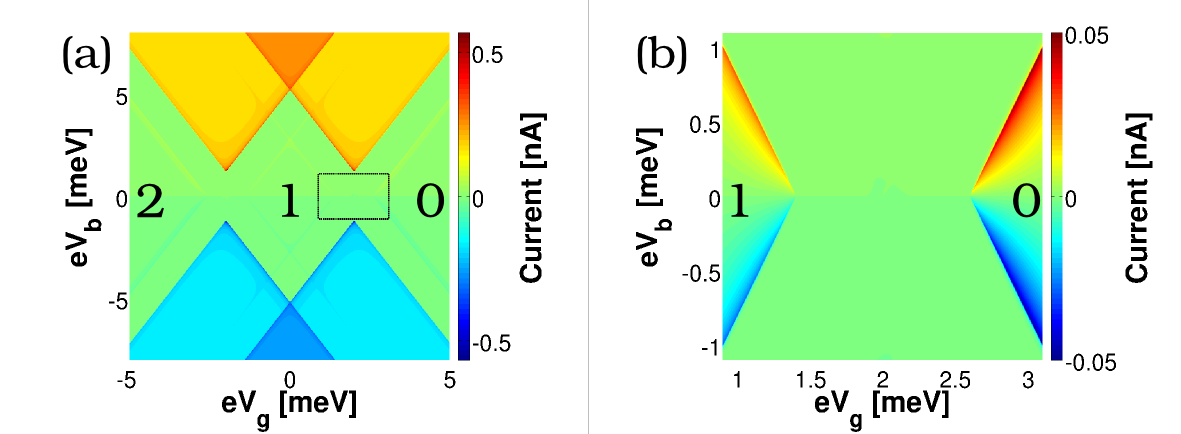}
\caption{(Color online) (a) Current voltage characteristics of a SD coupled to
superconducting leads. Parameters are
$k_B T = 0.3 \,\text{meV}$ and $|\Delta| = 0.6\,$meV,  $U =
4\,\text{meV} $, $\epsilon_d = -2 \,\text{meV}$, $e\Gamma = 0.001\,\text{meV}$.
(b) Subgap features
coming from thermally excited quasiparticles of the 0-1-particle transition,
highlighted as a dashed box in (a). }
\label{fig:SD_LT}
\end{figure}

The simplest example of a quantum dot system is the single level quantum dot
presented  in Eq.~(\ref{eq:SD}). Since only one level is
involved, we can do most calculations analytically and understand the basic
mechanism resulting from the  superconducting leads. 
 In Fig.~\ref{fig:SD_LT}  the stationary current is shown as a function of bias
and gate voltage for  
superconducting leads at $k_B T = 0.5 |\Delta| $.
 We observe the expected gap
\cite{Grove-Rasmussen2009} between the Coulomb
diamonds which is equal to  $4 |\Delta|/e$. The gap can be explained using Fig.
\ref{fig:trans_lines} and the corresponding Eqs.~(\ref{eq:s+})-(\ref{eq:dt-}).
One dashed line marks the gate voltage where $\Delta E_g = 0$. Along this line
the  conditions under which current is allowed to flow read: $eV_b/2 > |\Delta|$
for the S+, D- lines, and $eV_b/2 < -|\Delta|$ for the S-, D+ lines, opening a
bias window of $4|\Delta|/e$ where current is blocked for low temperatures $k_B
T \ll |\Delta|$. 
For higher temperatures of $k_B T \approx 0.5 |\Delta| $ we
observe small peaks in the  Coulomb blockade region (green area) which are due
to thermally excited quasiparticles; they correspond to the red lines in
Fig.~\ref{fig:trans_lines}.
 In Fig.~\ref{fig:single_level_E-N} we show the energy particle
number diagrams in the points (a)-(d), which lie on a vertical cut through
Fig.~\ref{fig:trans_lines} at
$\Delta E_g > |\Delta|$ which corresponds to a gate voltage $e V_g > 2.6 \,
$meV
in Fig.~\ref{fig:SD_LT}.
In Fig.~\ref{fig:single_level_E-N} (a) we depicted the $E_g$-$N$ diagram for a
cut
with the S+ resonance line, where the particle number on the system is
increased by a tunneling event at the source  and decreased at the drain
contact. For bias voltages smaller than the one at resonance (corresponding to
larger $eV_b$ as $e$ is the negative charge of an electron) the S+, D-
transitions remain open and current can flow. In
Fig.~(\ref{fig:single_level_E-N}) (b) the $E_g$-$N$ diagram at the resonance
line St+
is shown. In this case the bias voltage is not large enough to allow the
transitions S+ of Eq.~(\ref{eq:s+}). For low temperatures no quasi particle is
thermally excited  meaning that only transitions from $1\rightarrow 0$ are
energetically allowed (green arrows). For high
enough
temperatures, however, the particle number of the system can be increased by
tunneling events involving thermally excited quasiparticles opening the St+
transition.  By changing the sign of the
bias voltage the role of the source and the drain is inverted, explaining the
transition lines Dt+ and D+ (Fig.~\ref{fig:single_level_E-N}(c) and
\ref{fig:single_level_E-N}(d)).

Another interesting constellation of the energy level occurs in the region of
$0< \Delta E_g < |\Delta|$ (Fig.~\ref{fig:single_level_E-N} (e)), where in the
current-voltage characteristics the
thermal lines are vanishing. Transitions can only occur at the drain
contact, as the bias is not large enough to allow transitions at the source.
Hence, the system is in thermal equilibrium with the drain contact and the
occupation probabilities are related by the Boltzmann distribution:
\begin{equation}
 \frac{\rho_0}{\rho_1} = e^{\beta(\Delta E_g + e V_b/2 )},
\end{equation} 
in the limit of $\gamma \rightarrow 0$.

\subsection{The double quantum  dot}

We have seen that the theory can reproduce well known results for the SD and we
understood the properties of the thermal transitions in $E_g$-$N$ diagrams with
only one non degenerated level per particle number. In the
following we investigate a more advanced system, the double quantum dot, where
the many body spectrum gives rise to more than one
non degenerated level per particle number, so called excited system states. For
normal conducting leads the excitations cannot be seen for low bias voltages,
since transitions to the ground state are always possible, blocking transport
through the excitations. In the last subsection we have seen that for
superconducting leads the energy difference must be at least $|\Delta E_g| \geq
eV_b/2 - |\Delta|$ to have non thermal source and drain transitions. Hence,
we find situations where the transition to the ground state are energetically
not allowed  and transport occurs through excited system states.

We start with equally gated dots with the same on-site energies and on-site
Coulomb interactions, where it is possible to diagonalize the Hamiltonian
analytically \cite{Bulka2004,Hornberger2008}. 
In the second part, the case of independently coupled
dots is discussed, where the detuning of the two gate voltages influences
the level spacing of the energy spectrum. Thus, excited states can be observed
only in detuning ranges where the difference between the energy level of the 
excited state and its ground state is less than $2|\Delta|$.

\subsubsection{Equally gated dots}\label{sect:equally gated dots}

\begin{figure}
  \includegraphics[width = \columnwidth]{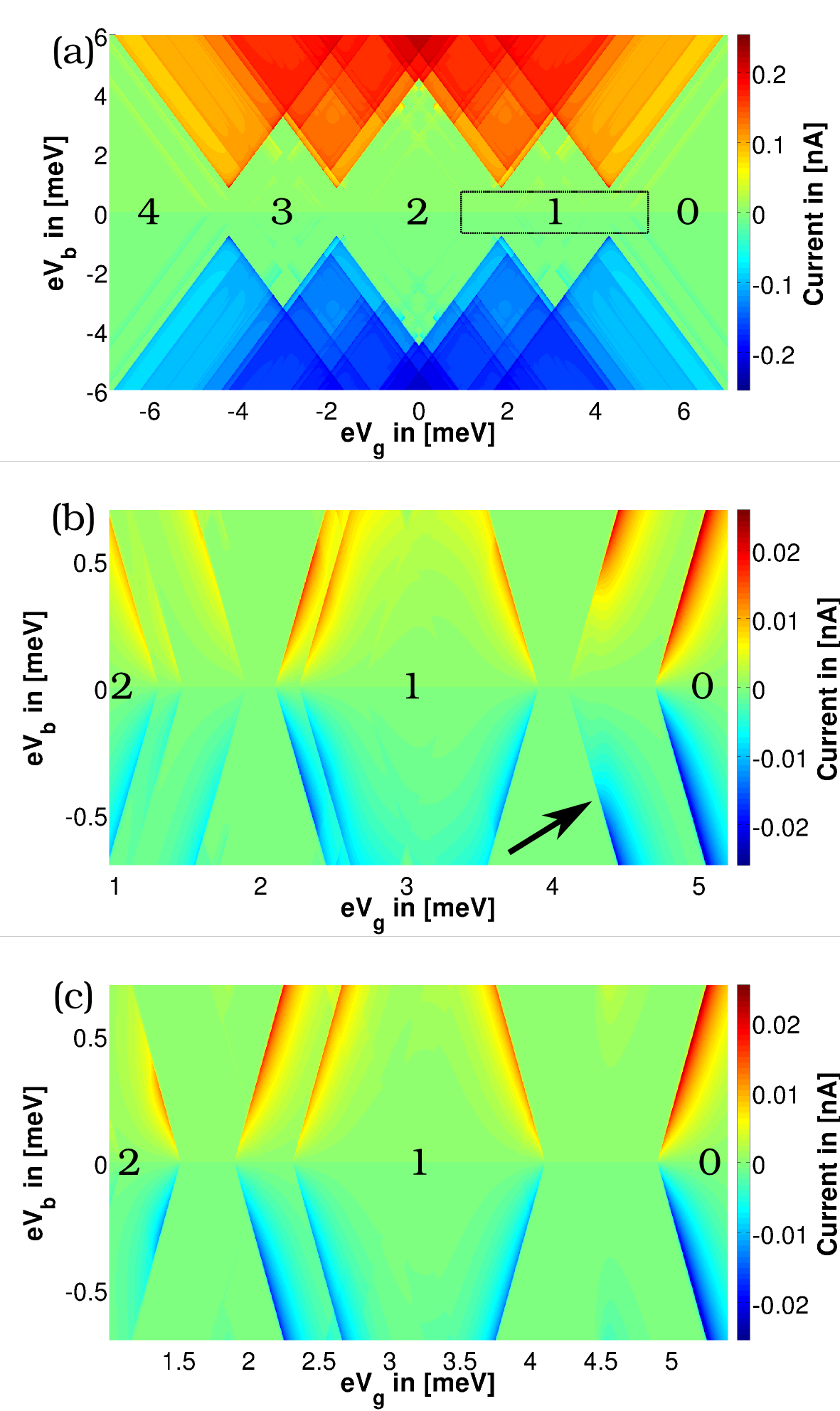}
\caption{(Color online) (a) Current voltage characteristics of an equally gated
DD in serial configuration  at $k_B T = 0.2\,\text{meV}$,
 $|\Delta| = 0.4\,\text{meV}$, $U = 4 \, \text{meV}$, $V = 2
\,\text{meV}$, $b= -0.3 \,\text{meV}$, and $e\Gamma = 0.001 \,\text{meV}$. (b)
I-V characteristics in the subgap region  corresponding to the dashed box in
(a). The distance between the 1-particle excited state and its ground state is
equal to the coupling strength $2|b|$  of the two dots. Moreover,
$2|b|<2|\Delta|$. The black arrow
marks the transition line coming from transport through the 1-particle excited
state.
 (c) I-V-characteristics in the subgap region, where we increased the coupling
between the two dots ($b= -0.5 \,\text{meV}$), leading to a
level spacing which is larger than
$2|\Delta|$, hence transport through the excited system state is not allowed
and the line disappears.
}
\label{fig:DD_eq_serial_current}
\end{figure}

\begin{figure}
\includegraphics[width=\columnwidth]{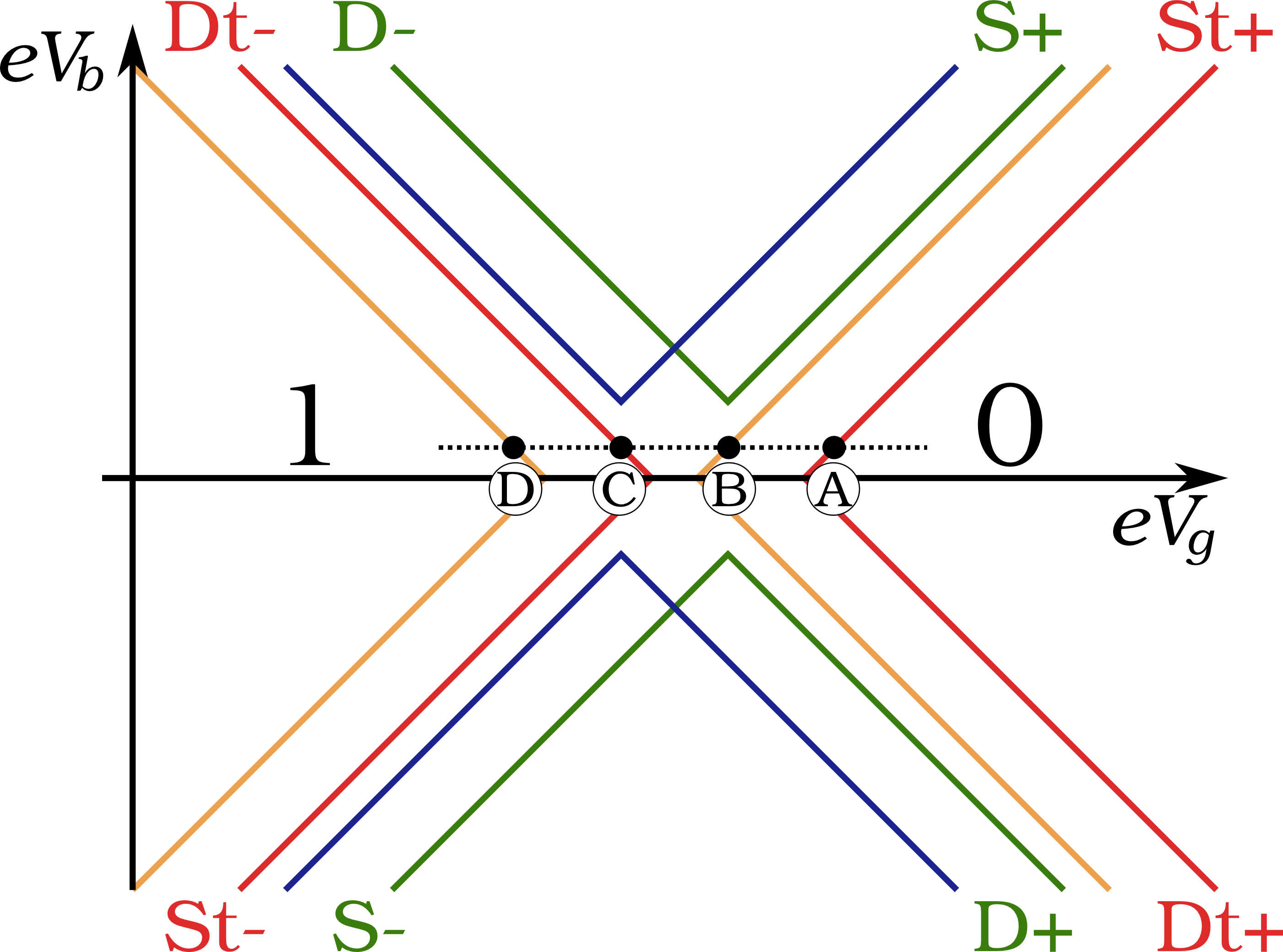}
\caption{(Color online) Sketch of the transition lines for the 0-1 particle
transition of an
equally gated DD. It shows two copies of Fig.~\ref{fig:trans_lines} where the
labeling of the blue (orange lines) is the same as for the green (red) lines.
The blue (orange) lines mark the transition lines corresponding to the 0-
particle ground state to 1-particle first excited state transition.
}
\label{fig:0_1_excited_line}
\end{figure}

\begin{figure}
\includegraphics[width = \columnwidth]{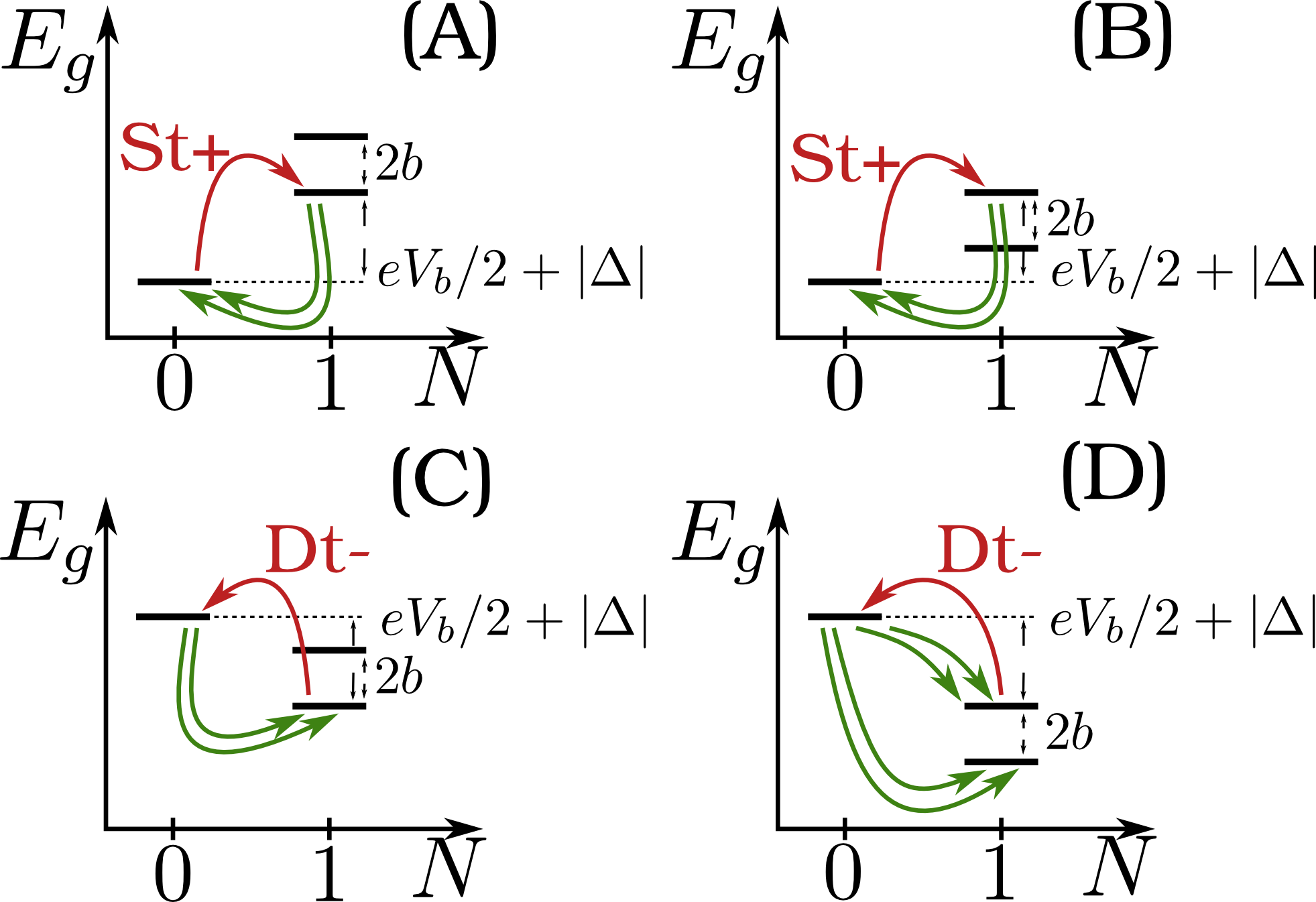}
\caption{(Color online) $E_g$-$N$ diagram corresponding to the points of
Fig.~\ref{fig:0_1_excited_line} where the dashed line cuts the transition lines
for
the case of an equally gated DD. In this case the distance between the
1-particle ground state to the 1-particle first excited state is equal to $2b<
2|\Delta|$, where $b$ is the tunneling strength between the two quantum dots.
(A) Point on the thermal line St+ of the ground state to ground state
transition. (B) Point on the thermal line St+ of the ground state to  first
excited state transition. (C) Point on the Dt- line of the ground state to
ground state transition. (D) Point on the Dt- line of the ground state to first
excited state transition; this line cannot be seen in the current voltage
characteristics, as the ground state to ground state transitions are open.
Hence, in the long time behavior the system will occupy the 1-particle ground
state blocking the current through the excited state. 
}
\label{fig:point_A_B_C_D}
\end{figure}

\begin{figure}
\centering
\includegraphics[width = 0.5
\columnwidth]{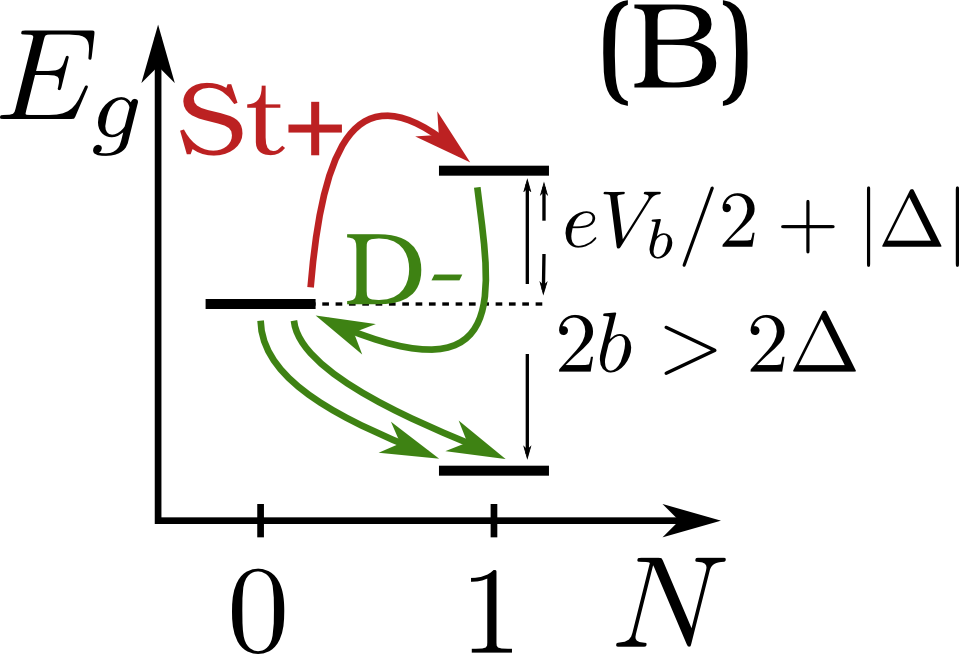}
\caption{(Color online) $E_g$-$N$ diagram of point (B) in
Fig.~\ref{fig:0_1_excited_line}, for
 a level spacing of the one particle energies  larger than $2b > 2\Delta$. In
contrast to Fig.~\ref{fig:point_A_B_C_D} the transition between the 0-particle
ground state and the 1-particle excited state is open, blocking the current.}
\label{fig:point_B_b>delta}
\end{figure}

For equally gated dots the on-site energies of the two sites are 
modulated with the same gate voltage. Hence, it is convenient to plot the
current as a function of the bias and the gate voltage as for the
SD. Fig.~\ref{fig:DD_eq_serial_current} shows the current of an equally gated DD
in
serial configuration. As for the SD we
observe Coulomb blockade and the gap of $4|\Delta|/e$ between the tips of the
diamonds. Transport carried by thermally excited quasiparticles is of
particular interest, as it allows one to observe transitions through excited system
states for low bias voltages, which are often diminished by the ground state
transitions in the normal conducting case.
In order to show some interesting phenomena resulting from the more
complex spectrum, we concentrate on the 0- to  1-particle
transition where three levels are involved. In the 1-particle spectrum, the
difference between the ground state and
the excited state is equal to $2|b|$, where $b<0$ is the tunneling strength
between the two dots. Meaning that by tuning the coupling between the two dots
it is possible to influence the level spacing. Fig.~\ref{fig:0_1_excited_line}
shows a sketch
of the transition lines expected for the $0-1$ transition for $|b|<|\Delta|$,
where the red (green) lines show the ground state to ground state transitions,
and the blue (orange) lines the ground state to first excited state transitions.
For a better understanding of the transport properties we cut the transitions
lines horizontally for a small bias voltage $eV_b/2 < |\Delta|$ in the Coulomb
blockade
region (points (A)-(D)), the corresponding $E_g$-$N$ diagrams are depicted in
Fig.~\ref{fig:point_A_B_C_D}. In point (A) the difference between the ground
states
is equal to $\Delta E_g = eV_b/2 +|\Delta|$ opening the thermal transition
St+ and current can flow. Following the dashed line to point (B), the 1 particle
states are shifted
down in energy until the St+ transition is allowed between the
0-particle ground state and the 1-particle excited state. Since $ |b| < \Delta$,
the 1-particle ground state is energetically not accessible and current can flow
through the excited state. We like to emphasize that the blocking of the ground
state transition is only valid as long as the distance between the two
1-particle levels is smaller than $2|\Delta|$. For larger distances the ground
state is energetically accessible, blocking the current through the excited
state, c.f. Fig.~\ref{fig:point_B_b>delta}.
 In point (C) $eV_g$ is further decreased, the Dt-
transition  between the ground states is opening, and current can flow. 
Point (D) shows the typical energy configuration in which current through the
excited state is blocked, even though the transition through the excited state
is energetically allowed. The reason for that is the 1-particle ground state
which can be populated, but transitions describing its depopulation are
energetically not allowed, leading to a blocking of the current in the
stationary limit.

To demonstrate the important role of the level spacing 
 we show the current voltage characteristics of an equally gated DD in the
subgap region in Figs.~\ref{fig:DD_eq_serial_current}(b) and  \ref{fig:DD_eq_serial_current}(c). In
(b) the spacing
of the 1-particle energy levels $|2b| < 2|\Delta|$, hence, the excited state can
be observed in the current (arrow in Fig.~\ref{fig:DD_eq_serial_current}). In
(c) we
increase the tunneling strength between
the two dots $2|b| > 2|\Delta|$ and the excited state line is vanishing, as
explained in Fig.~\ref{fig:point_B_b>delta}. As in the case for $2|b|<2|\Delta|$
the excited level is in resonance with the St+ transition, however, due to the
larger level spacing, the ground state transition opens and current is blocked.

\subsubsection{Independently gated dots}

\begin{figure}
\includegraphics[width = 0.8 \columnwidth]{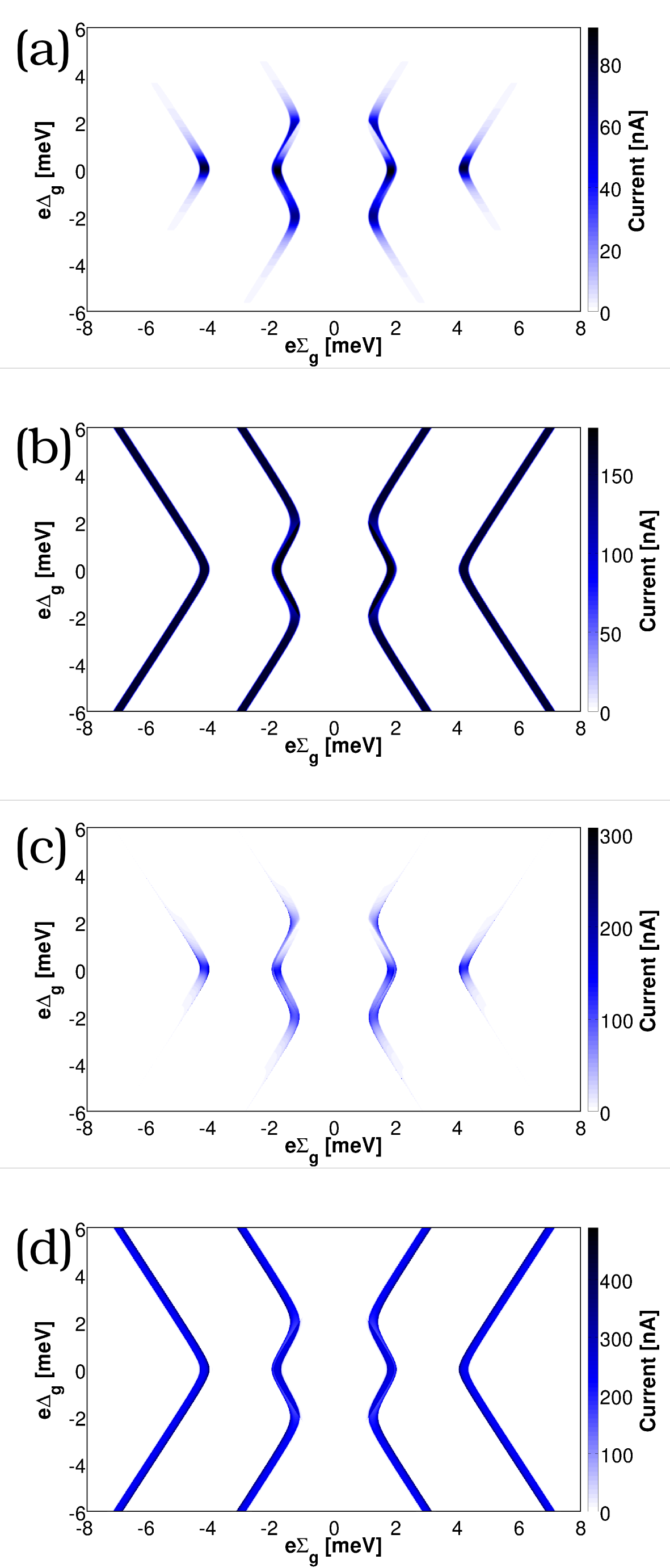}
\caption{(Color online) (a)-(b) Current voltage characteristics of a DD coupled
to normal conducting
leads in serial (a) and in parallel (b) configuration. We
fixed the bias voltage to $e V_b = 0.3\, $meV. 
(c)-(d)
Current voltage characteristics of a DD coupled to
superconducting
leads in serial (c) and in parallel (d) configuration. We
fixed the bias voltage to  $e V_b =  0.3\,\text{meV}+ 2|\Delta|$ in order to
obtain the same conditions as for the normal conducting case in (a)-(b). 
Parameters are: $T= 0.01\,\text{meV}$, $|\Delta| = 0.4
\,\text{meV}$, $e\Gamma = 0.001 \,\text{meV}$, $b= -0.2 \,\text{meV}$, $U = 4 \,
\text{meV}$ and $V= 2 \,\text{meV}$. 
}
\label{fig:DD_nc_sc}
\end{figure}

\begin{figure}
\centering
\includegraphics[width=\columnwidth]{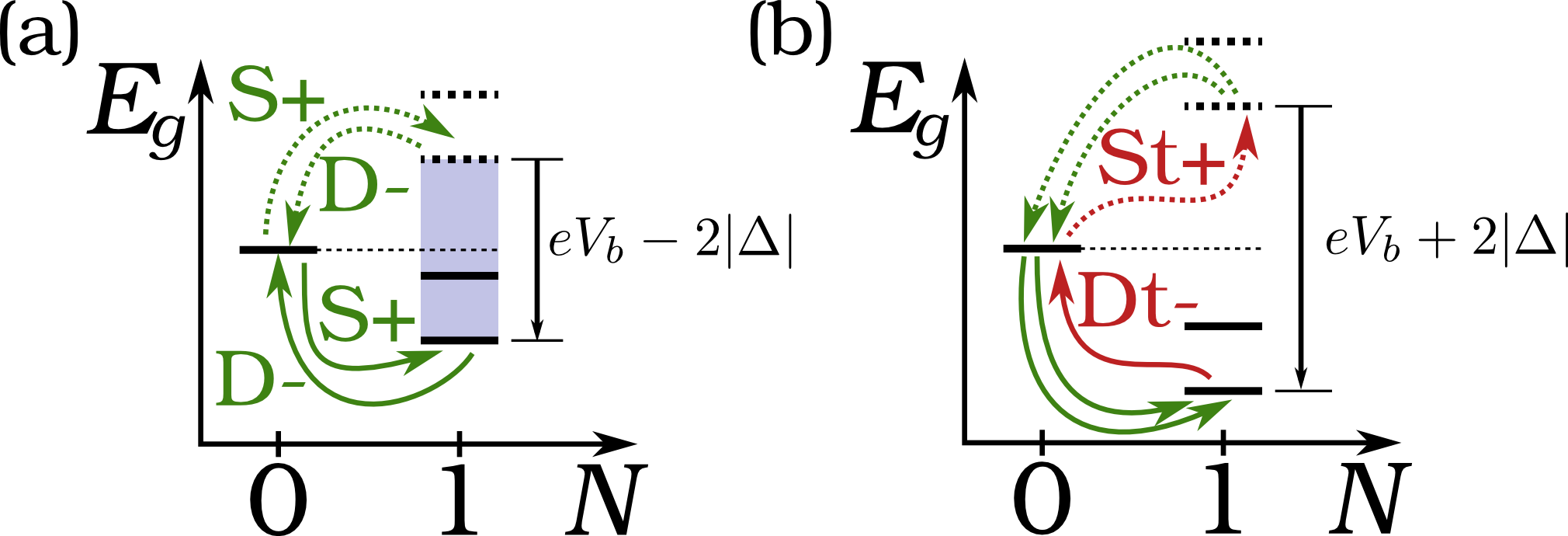}
\caption{(Color online) \textbf{(a)} $E_g$-$N$ diagram of the 0-1-particle
transition for $eV_b/2
>|\Delta|$.
In the 1-particle spectrum we plotted two situations which mark the borders
of the current step. The dashed levels mark the left border (for small
$\Sigma_g$) where the 1-particle levels lie above the 0-particle energy level. 
If the distance $\Delta E_g \leq eV_b/2  - |\Delta|$ current
can flow
through S+ and D- transitions. By lowering $e \Sigma_g$  the 1-particle energy
levels move down in the $E_g$-$N$ diagram, while the transitions remain open.
The
solid lines mark the right border of the current steps, as for levels lying
below the solid line the D- transition is closed and current is blocked. Thus,
the width of the current steps in the current voltage characteristics is:
$e \Delta \Sigma_g = eV_b -2 |\Delta|$. \textbf{(b)} $E_g$-$N$ diagram of the
0-1-particle transition involving thermal transitions. For the same arguments
as in (a), the distance between two thermal lines in the current voltage
characteristics is equal to $e \Delta \Sigma_g = eV_b + 2|\Delta|$.}
\label{fig:IDD_E-N-0-1}
\end{figure}

\begin{figure}
\centering
\includegraphics[width = 0.35\columnwidth]
{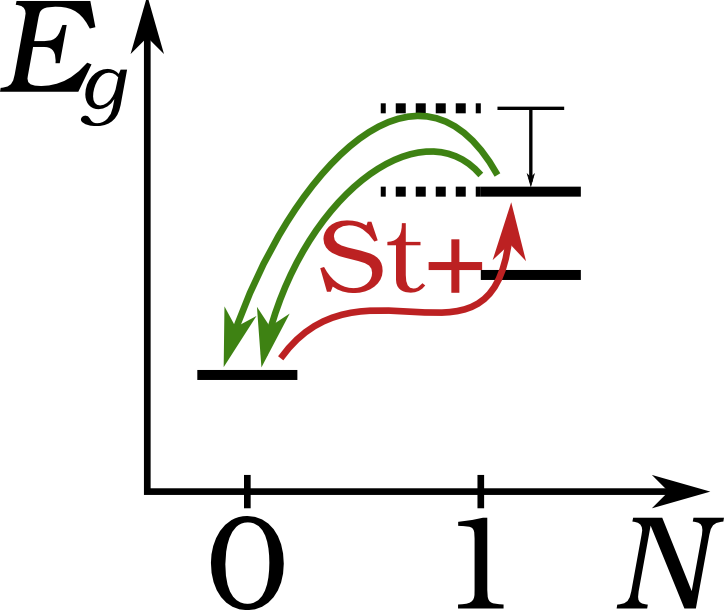}
\caption{(Color online)
$E_g$-$N$ diagram for the 0-1-particle transition. Transitions between the two
1-particle levels (dashed lines) and the 0-particle ground state are allowed
through the thermal St+ transition. Increasing the gate voltage the levels
move down in energy (solid lines) and the excited state transition can be
observed when the excited level aligns with the St+ transition. Hence,
 the distance of two neighboring thermal  transitions is
equal to the level spacing.  }
\label{fig:IDD_E-N-0-1_level_spacing}
\end{figure}

In the last paragraph we considered a DD with both dots coupled to the same
gate electrode. In most experiments, however, it is more convenient to couple
the dots independently, which leads to a 'honeycomb' shaped current voltage
characteristics \cite{VanderWiel2003}. For  symmetric on-site energies and
Coulomb repulsion it is possible to diagonalize the DD Hamiltonian of
Eq.~(\ref{eq:DD hamiltonian}) analytically. Gating
the dots independently destroys this symmetry,  an analytical
diagonalization is not possible, and one has to use numerical methods.  We plot
the current as a function of the detuning  $\Delta_g= V_{g}^1 - V_{g}^2$,
and the average of the two gate voltages $\Sigma_g = (V_{g}^1 + V_{g}^2)/2$.

The current voltage characteristic for  serial and parallel
configuration is depicted for the normal conducting
case in Fig.~\ref{fig:DD_nc_sc} (a)-(b)  and  for the
superconducting case in Fig.~\ref{fig:DD_nc_sc} (c)-(d).
 Comparing both configurations, we observe for the serial one a decrease in the
current for high detuning $\Delta_g$, while in the parallel configuration
current can be observed over the entire voltage range. This difference is a
consequence of the geometry of the set-up as the DD system remains unchanged. 
An increase of the detuning leads to a localization of the systems ground state 
at site 1 and transitions through site 2 are blocked.
Since in serial configuration the right lead is only coupled to site 2, the
localization of the wave function at site 1 leads to a decrease in the current.
In parallel configuration, however, both sites are coupled to both leads and the
ground state transition is always open.

The left and right border of the current steps are given by the  source
and drain lines, respectively. They follow, in complete analogy to the simplest
case, from energy conservation.  In Fig.~\ref{fig:IDD_E-N-0-1} (a)
we show the
$E_g$-$N$ diagram for the 0 to 1-particle  transition illustrating
two limits: the ground states are  (i) in resonance with the  S+ transition
(dashed line) and (ii) in resonance with the D- transition (solid line),
describing the left
and right borders of the current step in Fig.~\ref{fig:DD_nc_sc} (c-d). Starting
at the S+ resonance, the energy levels of the 1-particle spectrum are moving
down in energy by increasing the average gate voltage $\Sigma_g$. Both
transitions (S+ and D-) remain open as long as the ground state lies  in the
blue (shaded)
region. If the ground state lies below the solid line, the D- transition is
closed and current is blocked.
Hence, the width of the current steps in the current voltage characteristics in
Fig.~\ref{fig:DD_nc_sc} (c-d) is equal to the size of the blue (shaded) region
in Fig.~\ref{fig:IDD_E-N-0-1} (a), namely $e\Delta \Sigma_g = eV_b
-2|\Delta|$.
The same arguments hold for the distance of two corresponding thermal
transitions, as illustrated in Fig.~\ref{fig:IDD_E-N-0-1} (b) the
distance of two thermal lines is equal to $e\Delta \Sigma_g = eV_b + 2|\Delta|$.

As we can see in Fig.~\ref{fig:DD_nc_sc} there exists a one
to one correspondence of the transport conditions of the normal conducting to
the superconducting case which leads to the same shape of the current voltage
characteristics if $k_B T \ll |\Delta|$. 
Increasing the bias voltage by $2 |\Delta|$ compared to the normal conducting
case $eV_b^{\text{\tiny{SC}}} =e V_b^{\text{\tiny{NC}}} + 2 |\Delta|$ leads to
the same transport conditions. Although
the shape of the current steps in Figs.~\ref{fig:DD_nc_sc} (a-b) and
\ref{fig:DD_nc_sc} (c-d) look the same, they differ at the edges of the
current steps, as in the superconducting case the sharp peaks of the
quasiparticle density of states are reflected in the current.

\subsubsection{Thermal effects}\label{sect:thermal effects DD}

\begin{figure}
 
\includegraphics[width= \columnwidth]{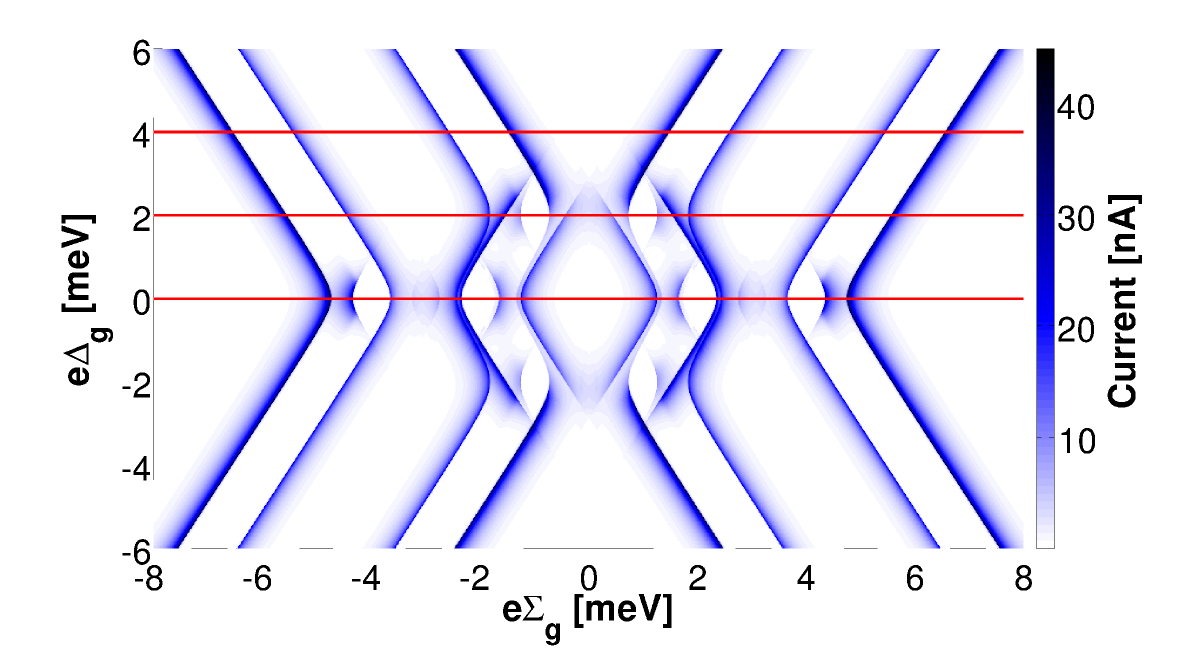}
\caption{(Color online) Current voltage characteristics of a DD in parallel
configuration for
bias $V_b < 2|\Delta|/e$. Since the bias voltage is not high enough current can
flow only due to thermally excited quasiparticles. The red lines correspond to
Fig.~\ref{fig:DD_E_diff} where the energy differences of the excited states
with respect to their ground state are plotted as a function of particle
number. The number of visible excited states is proportional to the number of
energy differences which are smaller than $2|\Delta|$ (red line in
Fig.~\ref{fig:DD_E_diff}). 
Parameters are: $T= 0.01\,\text{meV}$, $eV_b = 0.3
\,\text{meV}$ $|\Delta| = 0.4
\,\text{meV}$, $e\Gamma = 0.001 \,\text{meV}$, $b= -0.2 \,\text{meV}$, $U = 4 \,
\text{meV}$ and $V= 2 \,\text{meV}$. 
}
\label{fig:DD_parallel_subgap}
\end{figure}

\begin{figure}
 \centering
\includegraphics[width= 0.7 \columnwidth]{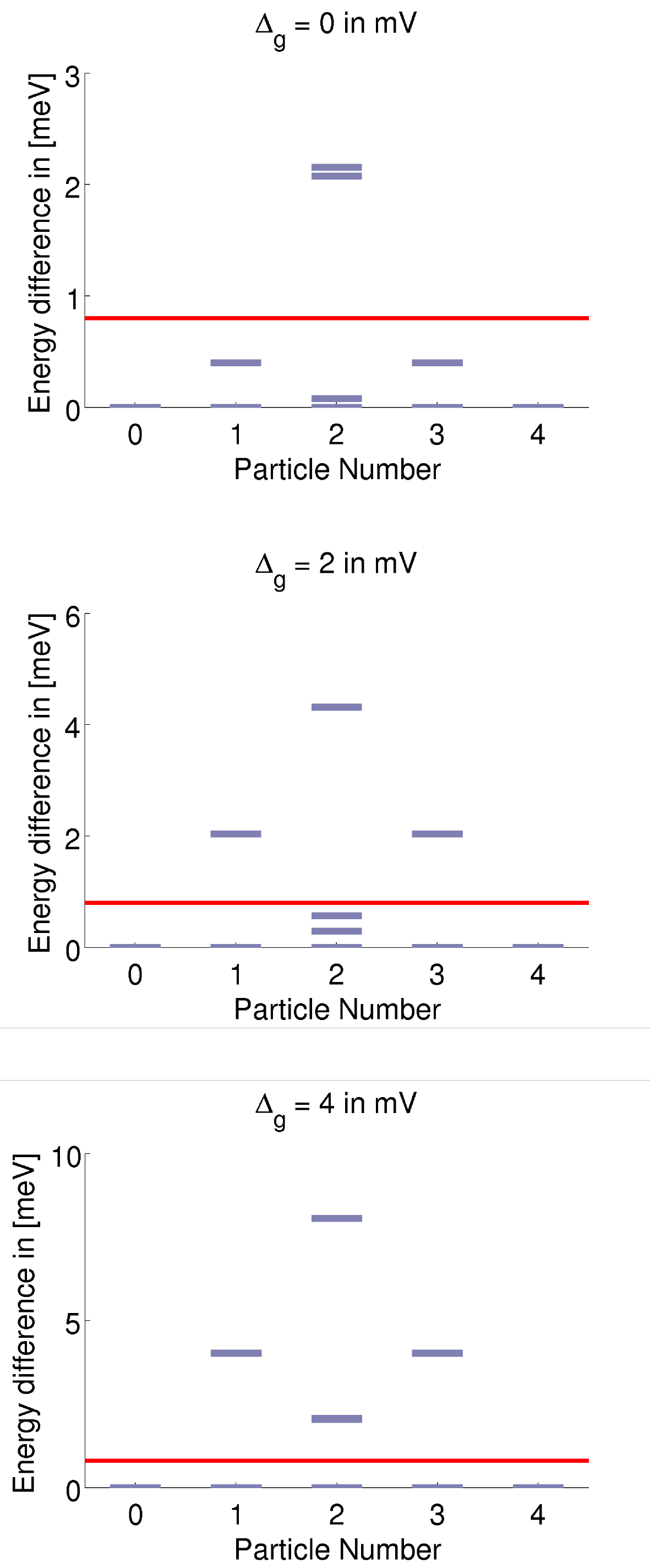}
\caption{(Color online) Plot of the energy differences of the excited system
states with
respect to their ground state as a function of particle number.
If the energy difference is smaller than $2|\Delta|$, transitions through these
excited states can be seen in the current voltage characteristics. The threshold
of $2|\Delta|$ is marked as a red  horizontal line.  We depicted the plots for
three situations differing in the detuning $\Delta_g$. The three cases are
marked as horizontal lines in Fig.~\ref{fig:DD_parallel_subgap}.}
\label{fig:DD_E_diff}
\end{figure}

We have seen that the  shape of the stability diagram can be
explained using energy conservation, in complete analogy to the simplest case.
In this section we  discuss the case for  small bias voltages $eV_b/2
< |\Delta| $, where current can flow due to thermally excited quasiparticles
exclusively. As already observed above, thermally excited
quasiparticles do not produce steps in the current voltage characteristics 
rather they appear as small peaks. This can be used to resolve transitions
through excited system states whose energy difference to the ground state is
less than $2 |\Delta|$. By detuning the gate voltages of the two sites of the
DD we can change the level spacing of the systems eigenenergies; hence,
 the excited states are only observed in a certain detuning range. 
To analyze transitions through excited system states, c.f.
Fig.~\ref{fig:DD_parallel_subgap}, we choose the parallel configuration to rule
out
the geometrical effect also leading to a decrease of the current for high
detuning.
 If a line corresponding to an excited state disappears for higher detuning
$\Delta_g$, we conclude that the  energy difference to its ground state is
larger than $2|\Delta|$. In Fig.~\ref{fig:DD_E_diff} we plotted the energy
differences of the excited states with respect to their ground state for
different values of
the detuning $\Delta_g$, which are marked as red lines in
Fig.~\ref{fig:DD_parallel_subgap}.  Counting the number of levels lying under
the
red line in Fig.~\ref{fig:DD_E_diff} gives information about the number of
visible excited lines. For instance, consider the case of $\Delta_g = 0$ in
Fig.~\ref{fig:DD_E_diff}. Following the red line from small to high $\Sigma_g$
in Fig.~\ref{fig:DD_parallel_subgap}, we cross the $0$-$1$ particle transitions
and observe three lines: two corresponding to the ground state, and one line
in between corresponds to a transition through the 1-particle excited state. 
The distance between the leftmost ground state transition line and the excited
line determines the level spacing of the one particle spectrum, see
\ref{fig:IDD_E-N-0-1_level_spacing}.
In the 2-particle spectrum the energy difference of one excited state lies under
the red line. Hence we should see two  lines coming from excited system states,
namely the transition
between the  1-particle ground state and the 2-particle excited state, 
and transitions between the 2-particle ground state and the 1-particle excited
state.
Along the horizontal cut at $\Delta_g = 2$ in Fig.~\ref{fig:DD_parallel_subgap},
excited states can only be observed for the 1-2 particle and the 2-3 particle
transition. This is in agreement with Fig.~\ref{fig:DD_E_diff}, where only in
the 2 particle subspace energy differences lie under the threshold of
$2|\Delta|$.
For higher detuning, e.g. $\Delta_g=4$, no excited states can be seen, as the
detuning increases the level spacing, and all energy differences are larger than
$2 |\Delta|$ Fig.~\ref{fig:DD_E_diff}.


\subsection{The N-QD-S junction}

\begin{figure}
\includegraphics[width=\columnwidth]{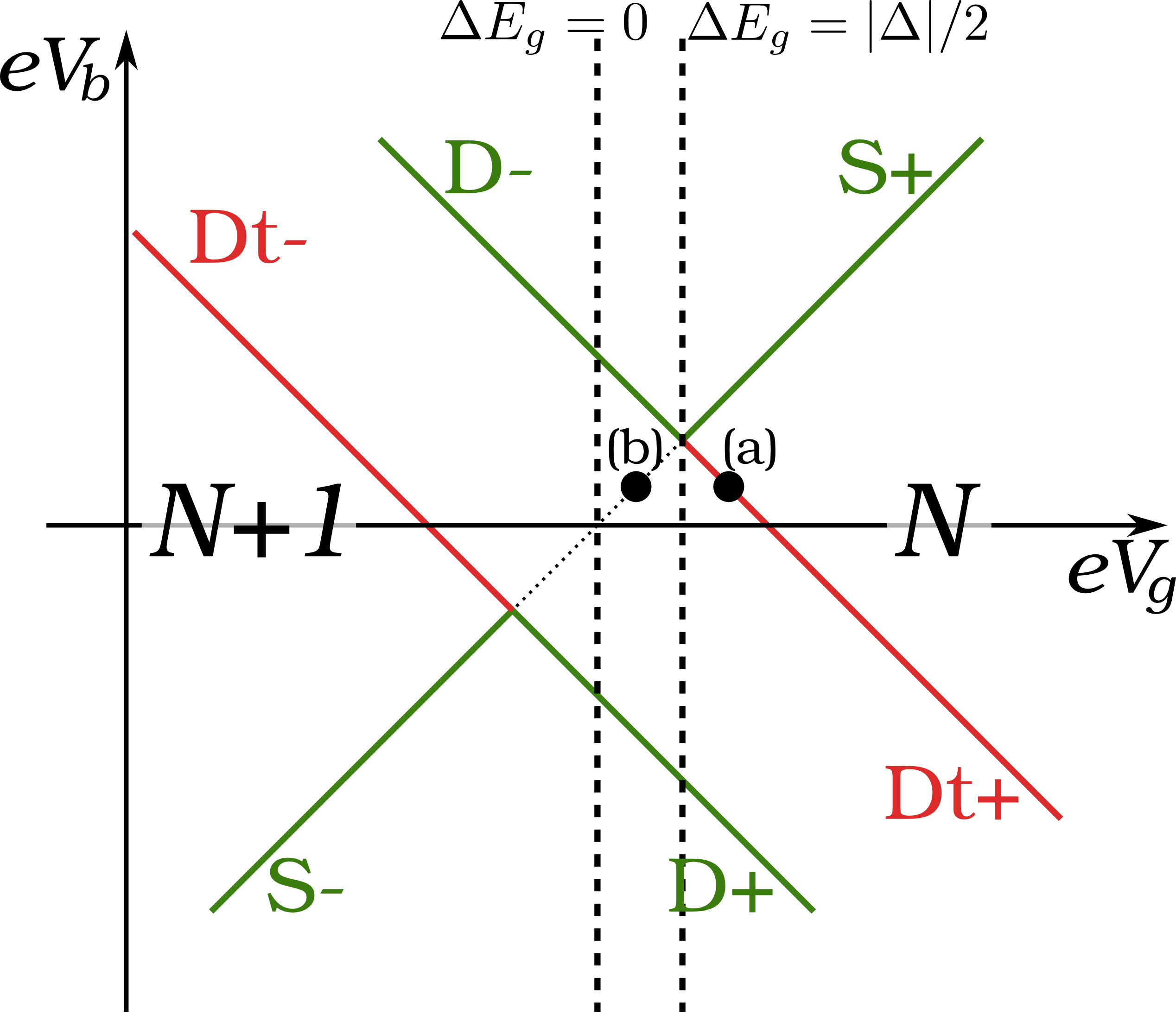}
\caption{(Color online) Sketch of the transition line of a QD coupled to a
normal conducting (source) and a superconducting lead (drain). The difference
to the S-QD-S system is that only the drain lines split due to the
superconducting gap, the S+ and S- lines are described by the same equation. In
this
case a gap equal to $|\Delta|$ is opening, and the triangles are shifted apart.
Thermal lines can be observed only for the drain.}
\label{fig:trans_lines_n-qd-s}
\end{figure}
\begin{figure}
\includegraphics[width=\columnwidth]{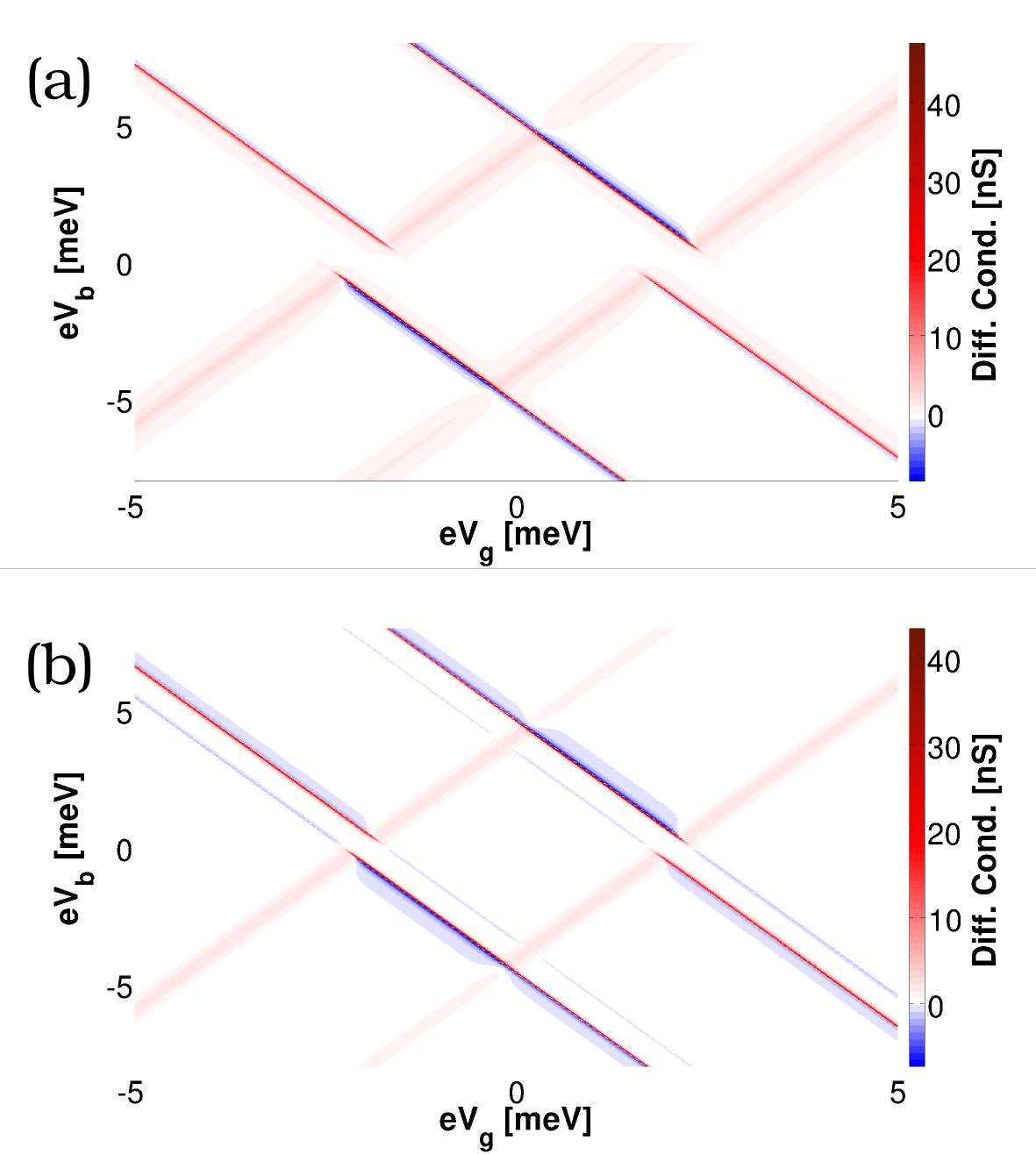}
\caption{(Color online) Differential Conductance of a SD coupled to a normal
conducting (source) and to a superconducting lead (drain) (N-QD-S system).
The coupling to the lead is $e\Gamma = 0.01\,$meV. (a)
Superconducting gap of $|\Delta| = 0.6\,$meV and temperature $k_B T = 0.1$meV.
No thermal lines in the subgap region are visible. (b) The same temperature $k_B
T = 0.1$meV,
but for smaller gap $|\Delta| = 0.3$meV; quasiparticles get thermally excited
across the gap leading to transport in the Coulomb blockade region. 
Parameters are $U= 4 \,\text{meV}$ and $\epsilon_d = -2
\,\text{meV}$. 
 }
\label{fig:n-qd-s_diff_cond}
\end{figure}
\begin{figure}
\includegraphics[width=0.5\columnwidth]{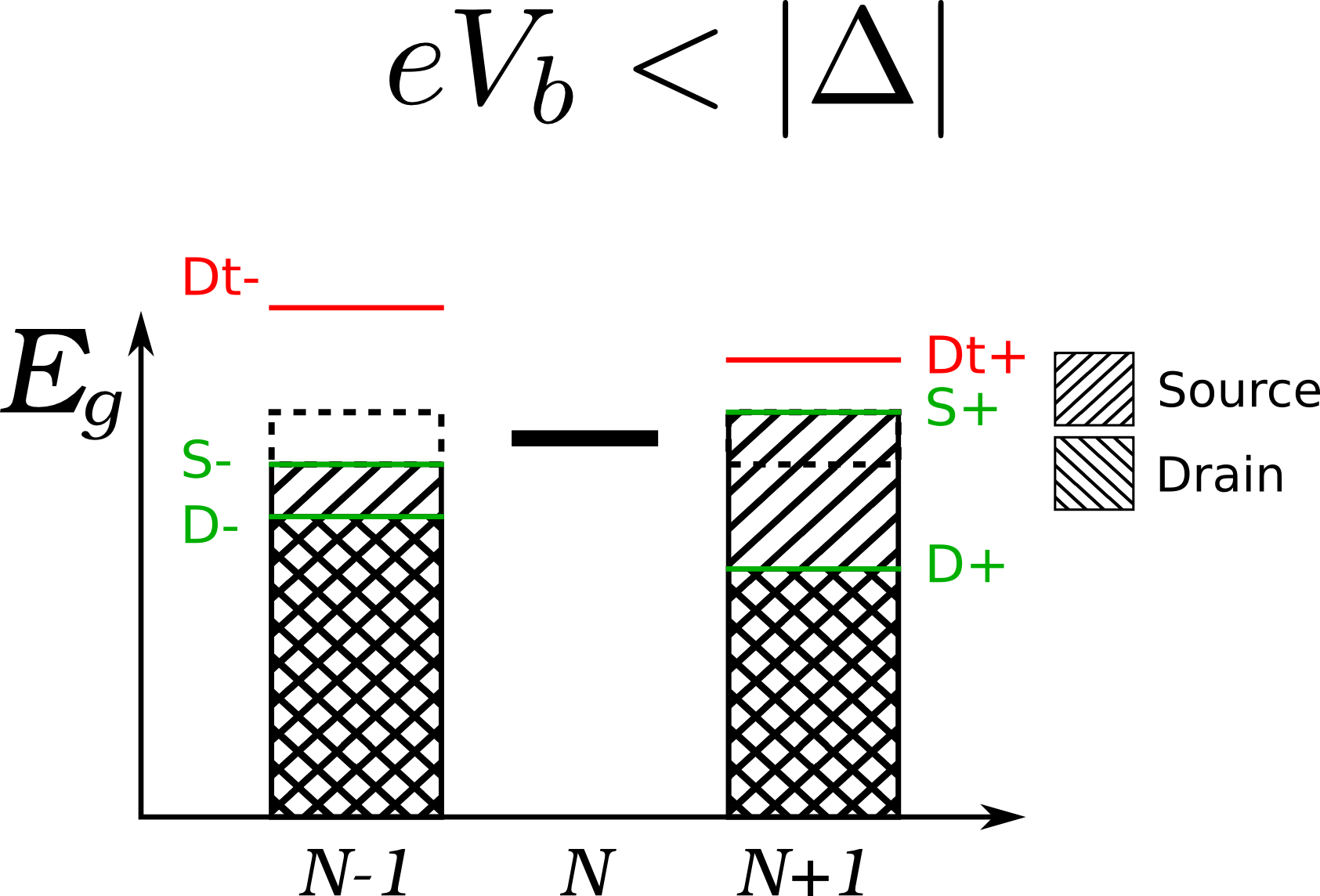}
\caption{(Color online) Visualization of the transport conditions for a N-QD-S
system with $eV_b/2 < |\Delta|$, where the source is a normal and the drain a
superconducting lead. They follow from Eqs.~(\ref{eq:s+})-(\ref{eq:dt-}) by
setting $|\Delta| = 0$ in the equations corresponding to the source lead. }
\label{fig:scheme_n-qd-s}
\end{figure}
\begin{figure}
\includegraphics[width=\columnwidth]{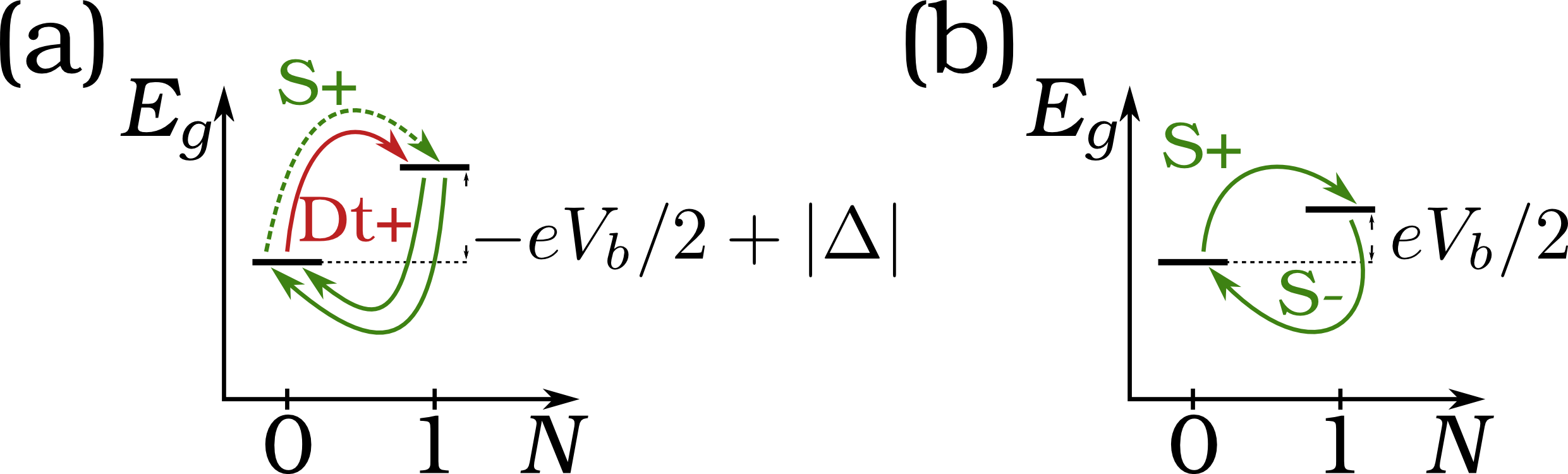}
\caption{$E_g$-$N$ diagrams corresponding to points (a) and (b) of
Fig.~\ref{fig:trans_lines_n-qd-s}. (a) We see a positive current in the subgap
region, which comes only due to the thermal smearing of the S+ transition. (b)
The line connecting the S+ and the S- transition line in the Coulomb blockade
region the system is in thermal equilibrium with the source contact. }
\label{fig:E-N-diagram_n-qd-s}
\end{figure}

We close this paper by investigating a so called N-QD-S hybrid
system, where a quantum dot system is coupled to a normal and to a
superconducting
lead, giving a possible explanation for the subgap features in Ref.
\onlinecite{Dirks2009}.
 In the experiment of Ref. \onlinecite{Dirks2009} a carbon
nanotube was contacted  to two normal conducting leads and to a superconducting
finger in between.
The differential conductance between the superconducting finger and a normal
lead is measured, realizing a N-QD-S hybrid system. It is possible to apply a
bias voltage across the entire tube as well as between the superconductor and
a normal conducting lead.
 The stability diagram in Fig.~2 (a) in Ref. \onlinecite{Dirks2009}, with no
bias applied over the entire tube, reveals
the typical Coulomb diamond pattern resulting from quasiparticle tunneling with
no subgap features. By applying a bias voltage $V_{SD}$ over the entire tube,
the gap in the stability diagram gets smaller with respect to the unbiased case 
and conductance lines can be seen in the Coulomb blockade region, c.f. Fig.~3
(a) of Ref. \onlinecite{Dirks2009}.
The reduction of the gap in the stability diagram is proportional to the
applied bias voltage of  approximately $eV_{SD} \approx |\Delta|/2$, and 
 is related to an effective reduction of the superconducting gap. For a smaller
gap quasiparticles can get thermally excited across the gap leading to subgap
transport in complete analogy to the S-QD-S case discussed above.

We can model the N-QD-S system by setting $|\Delta_S| = 0$ for the normal
conducting lead (source) in the master equation; the drain contact remains
superconducting $|\Delta_D| = |\Delta|$.
Hence, the transport conditions change slightly and can be summarized in the
scheme of Fig.~\ref{fig:scheme_n-qd-s}.
 In Fig.~\ref{fig:trans_lines_n-qd-s}
we schematically sketched the expected transition lines  for a N-QD-S
hybrid structure. In Fig.~\ref{fig:E-N-diagram_n-qd-s} we analyzed the two most
important cases, marked as points (a) and (b) in
Fig.~\ref{fig:trans_lines_n-qd-s}. Point (a) shows a paradoxical situation as
the
particle number of the system seems to be increased only at the drain contact,
which would lead to a negative current at positive bias. However, if the two
contacts have the same temperature, the thermal broadening of the S+ line gives 
a small contribution in the transition rates (dashed green arrow in
Fig.~\ref{fig:E-N-diagram_n-qd-s} (a)) making the current positive. The
situation in
(b) shows again the system being in thermal equilibrium with the source contact.

 We can see that the lines with negative slope (drain lines) give a
finite current in the Coulomb blockade region as observed in 
Fig.~3 (b) in the experiments. Thus, we claim that the subgap features
observed in the experiments possibly are transitions involving thermally excited
quasiparticles which are allowed due to the reduction of the superconducting
gap. This argument is supported by the observation that for diamonds where the
gap has the same size as before (edges of the stability diagram), no subgap
lines can be observed. In Fig.~\ref{fig:n-qd-s_diff_cond} we show two
$dI/dV-$ characteristic of a N-QD-S system corresponding to 
 different superconducting gaps with the same temperature ($k_B T = 0.1$meV) in
both cases. In (b) the superconducting gap ($|\Delta| = 0.3$meV) is only
half of the gap in (a) ($k_B T = 0.6$meV).  By reducing the gap, the 
temperature becomes  large enough to excite quasiparticles across the gap,
leading to conductance peaks in the
Coulomb blockade region, as observed in the experiments
 However, a
more complex modeling  of the multi-terminal system is required to understand
the experiments in all details.

\section{Conclusion}

In this work we developed a transport theory for nanostructures coupled to
superconducting leads up to second order in the tunneling Hamiltonian. We used
 the Bogoliubov transformation to describe the electrons in the superconductors
as Cooper pairs and Bogoliubov quasiparticle excitations, whereby we modified
the Bogoliubov transformation in a number conserving
way \cite{Josephson1962,Bardeen1962}, introducing Cooper pair creation and
annihilation operators explicitly. We showed the predictions of the theory on
two examples, the well known single level quantum dot, and the double quantum
dot. The
characteristic gap in the Coulomb diamonds, proportional to the superconducting
gap, as well as negative differential conductance was observed in both cases.
Further, we  considered the double quantum dot in serial as well as in parallel
configuration, see Fig.~\ref{fig:DD_configuration}, coupling the dots 
to the same as well as to two separate gate electrodes.

We systematically analyzed the stability diagrams, extending the scheme of
Ref. \cite{Donarini2010} for superconducting leads. We found that transport
through excited system states occurs even for low bias voltages using thermally
excited quasiparticles, leading to zero bias  peaks in the conductance.
Transitions through excited states can be observed if
transitions through the ground state are energetically not allowed, namely if
the distance between the energy levels of the excited state and the ground state
is smaller than $2|\Delta|$. This effect can be seen in the the current voltage
characteristics  of an independently gated double quantum dot  in parallel
configuration without tuning parameters of the system, since the level spacing
changes with the detuning $\Delta_g$ of the gate voltages. Hence the excited
states can be seen only in certain detuning windows. 
Finally, we analysed the case where a quantum dot is coupled to a normal and a
superconducting lead, giving a possible explanation for the subgap features of
Ref. \onlinecite{Dirks2009} in terms of  transport involving thermally excited
quasiparticles.

We conclude with the observation that thermally excited quasiparticles can lead
to a finite current in the Coulomb blockade region. Besides the well known
thermal transitions through the ground states, transitions through excited
system states must be taken into account  as they are an additional source of
zero bias peaks in the conductance. For a
better comparison with experiments the theory can be used to investigate more
realistic systems such as carbon nanotube quantum double dots. Specifically, the
current voltage spectroscopy in the low bias regime can be used to learn
something about the spectrum of the set-up.
Within our approach it is not possible to capture Josephson
current and Andreev reflections as they are higher order processes. Yet, in the
weak coupling regime lowest order quasiparticle transport gives not
only the basic structure of the Coulomb diamonds but also the
dominant subgap feature, i.e. thermally activated conductance peaks
associated to quasiparticle transport.
In order to observe the Josephson effect and Andreev reflections, the theory
must be extended to higher order perturbation theory
\cite{Governale2008,Martin-Rodero2011}.

\begin{acknowledgments}
We acknowledge financial support through DFG Program Nos. SFB631.
\end{acknowledgments}

\begin{appendix}

\section{Properties of the Cooper pair operators}
\label{app:bcs GS CPO}

In the microscopic description of superconductive tunneling it is
necessary to know the analytical form of the Cooper pair operators.
However, a microscopic discussion of the Cooper pair
operators and their influence on the transport properties of the
hybrid superconductor-quantum dot junction is rather rare in the
literature.
In this section we show the connection between the Cooper pair operators and
ground state of the particle number conserving lead Hamiltonian.
Starting from the definition of Eq.~(\ref{eq:definition_s}), we can formally
define the Cooper pair annihilation operator \cite{Schrieffer} as
\begin{equation}\label{eq:explicit_def_s}
\s= \sum_{M = 0 }^{\infty} \sum_{\{n_{k\si}\}} \ket{\{n_{k\si}\},2M}
\bra{\{n_{k\si}\},2M +2} ,
\end{equation}
where $\{n_{k\si}\} = \{n_{k_1\si_1},n_{k_2\si_2},\dots\}$ is a set of
quasiparticle 
occupation numbers. It follows that
\begin{equation}
\s\sdag = 1, 
\end{equation}
where we used 
\begin{equation}
 1 = \sum_{M =0 }^{\infty}\sum_{\{n_{k\si}\}} \ketbra{\{n_{k\si}\},M}.
\end{equation}
In the full Hilbert space the Cooper pair creation and
annihilation operators do not commute
\begin{equation}
 \big[\s,\sdag\big] = \hat{\mathcal{P}}_0,
\end{equation} 
where $\hat{\mathcal{P}_0}$ is the projector to states with zero Cooper pairs:
\begin{equation}
\hat{\mathcal{P}}_{0} = \sum_{\{n_{k\si}\}}
\ket{\{n_{k\si}\},0} \bra{ \{n_{k\si}\},0}.
\end{equation}
Using that $\hat N \ket{\{n_{k\si}\}, M } = (
N_{\{n_{k\si}\}}^{\text{\tiny{QP}}} + M) \ket{\{n_{k\si}\}, M } $, with
$N_{\{n_{k\si}\}}^{\text{\tiny{QP}}}$ being the number of quasiparticles in the
string ${n_{k\si}}$, one obtains:
\begin{equation}
\begin{split}
\big[ \hat N, \s \big] &= -2\s,\\
\big[ \hat N, \sdag \big] &= 2\sdag.
\end{split}
\end{equation}

 \section{Rates}
\label{app:rates}

\subsection{Normal rates}
In the stationary limit, ${\tau \rightarrow \infty}$, the normal rates
read:
\begin{widetext}

\begin{equation}\label{eq:gamma+_n->n+1}
 \begin{split}
&
\bigl( \Gamma^{+}_{n m m'n'}\bigr)_{\eta}^{N\rightarrow N+1} 
= \lim_{\tau \rightarrow \infty} \biggl( \frac{1}{\hbar}  \biggr)^2  \sum_{
k \si \alpha \alpha'} 
t_{\eta \alpha \si} t^*_{\eta \alpha' \si} 
\bra{n} \dan{\alpha \si} \ket{m}\bra{m'} \dddag{\alpha' \si} \ket{n'} \\
&
\int_0^{\tau} dt_2 ~e^{\ih E_{n'm'} t_2} 
 \biggl[ 
|u_{\eta k}|^2 f^+(E_{\eta k}) e^{+\ih (E_{\eta k} + \mu_{\eta})t_2} + 
|v_{\eta k}|^2 f^-(E_{\eta k}) e^{-\ih(E_{\eta k} - \mu_{\eta})t_2}
\biggr],
 \end{split}
\end{equation}
\begin{equation}\label{eq:gamma+_n->n-1}
\begin{split}
&
\bigl( \Gamma^{+}_{n m m'n'}\bigr)_{\eta}^{N\rightarrow N-1}  = 
\lim_{\tau \rightarrow \infty} 
\biggl ( \frac{1}{\hbar} \biggr)^2 \sum_{ k \si \alpha \alpha'}
t_{\eta \alpha' \si} t^*_{\eta \alpha \si} 
\bra{n} \dddag{\alpha \si} \ket{m}\bra{m'} \dan{\alpha' \si} \ket{n'} \\
&
\int_0^{\tau} dt_2~ e^{\ih E_{n'm'} t_2} 
 \biggl[ 
|u_{\eta k}|^2 f^-(E_{\eta k}) e^{-\ih (E_{\eta k}+ \mu_{\eta})t_2}
+ |v_{\eta k}|^2 f^+(E_{\eta k}) e^{+\ih(E_{\eta k} - \mu_{\eta})t_2}
\biggr].
\end{split}
\end{equation}
\end{widetext}

In the following we will show how to write the rates in
Eqs.~(\ref{eq:gamma+_n->n+1}) and (\ref{eq:gamma+_n->n-1}) in terms of an
integral
over quasiparticle energies $E_{\eta k}$. Neglecting the lead index $\eta$,
the energetic part of Eq.~(\ref{eq:gamma+_n->n+1}) is proportional to
\begin{equation}\label{eq:energetic_part_Gamma}
\begin{split}
\bigl( \Gamma^{+}_{n m m'n'}\bigr)^{N\rightarrow N+1}  \propto 
\sum_k\big(|u_k|^2  F_1(E_k) +  |v_k|^2  F_2(E_k) \big)
\end{split}
\end{equation} where we defined
\begin{equation}
\begin{split}
&F_1(E_k) = f^+(E_k) e^{\ih(E_k + \omega)t_2} ,\\
&F_2(E_k) =f^-(E_k) e^{-\ih(E_k - \omega)t_2},
\end{split}
\end{equation} 
with $\omega = E_{n'm'} + \mu_{\eta}$. Recalling the definition of $u_k$ and
$v_k$, c.f. Eqs.~(\ref{eq:def_uk}) and (\ref{eq:def_vk}), we see that
\begin{equation}\label{eq:connection_u_v}
\begin{split}
&|u_k(-\xi_k)| = |v_k(\xi_k)|. \\
\end{split}
\end{equation}  Writing the sum as $\sum_k \to
\int_{-\infty}^{\infty}d\xi_k \rho_N$, and exploiting
Eqs.~(\ref{eq:connection_u_v}) and (\ref{eq:u2+v2}) we are able to to write
 Eq.~(\ref{eq:energetic_part_Gamma}) as: 
\begin{equation}
\begin{split}
&\int_0^\infty d\xi_k \big( F_1(E_k) + F_2(E_k)  \big) .
\end{split}
\end{equation}
 Changing the integration variable from $\xi_k>0 \to E_k$
we obtain\begin{equation}
\begin{split}
\int_{|\Delta|}^{\infty}dE~D(E)\big( F_1(E) + F_2(E) \big),
\end{split}
\end{equation}
where we defined the superconducting density of states as
$D(E)=\rho_N\,\text{Re}\big(\frac{|E|}{\sqrt{E^2 + |\Delta|^2}})$. Due to the
definition of the density of states with the real part, we can extend the
integral to zero, and use that $F_2(-E) = F_1(E))$ to obtain
\begin{equation}
\begin{split}
\int_{-\infty}^{\infty}dE\,D(E) F_1(E).
\end{split}
\end{equation}

\subsection{Renormalization of the rates}\label{app:renormalization}

\begin{figure}
 \includegraphics[width= 0.6 \columnwidth]{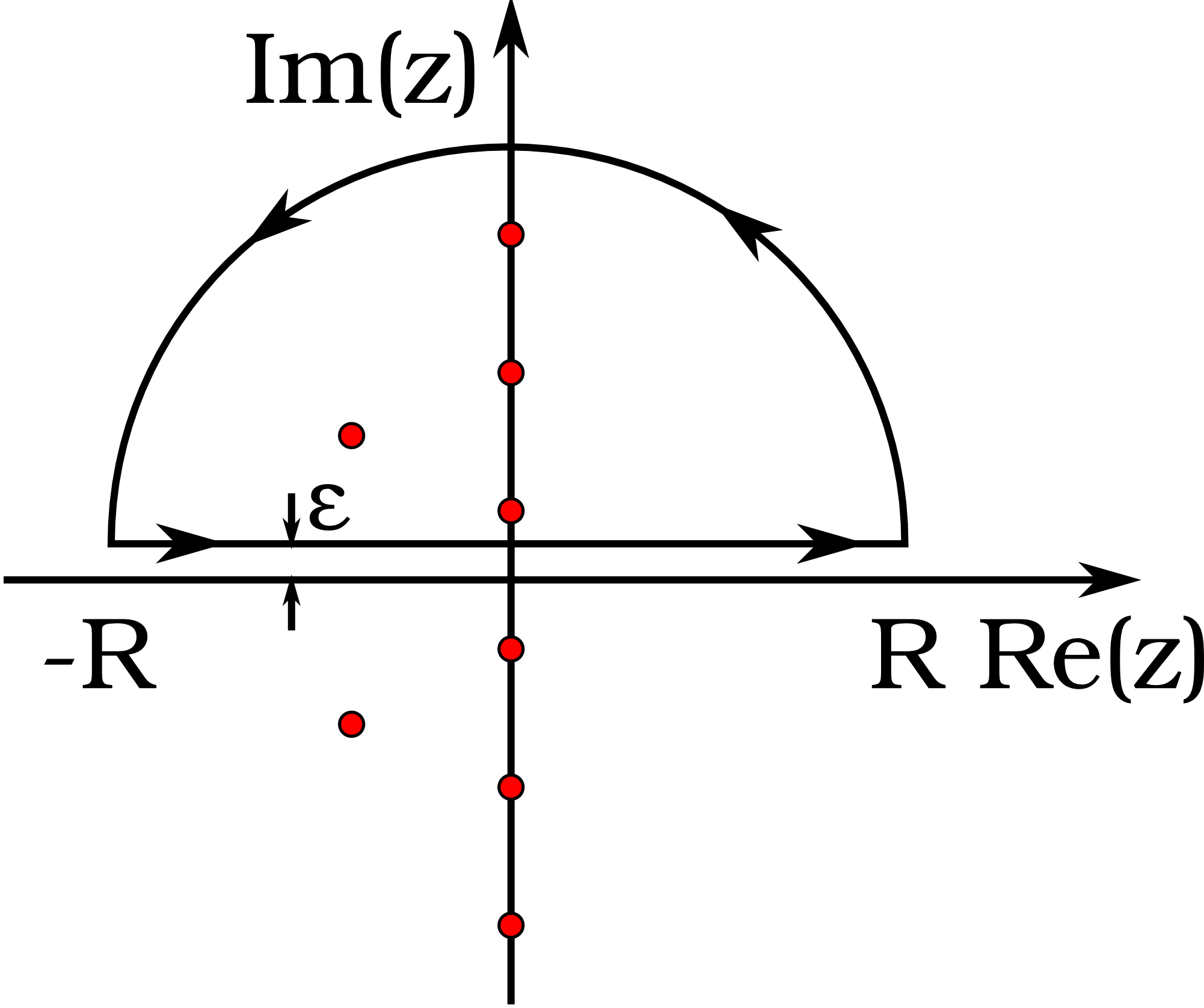}
 \caption{Contour in the complex plane used to integrate
Eq.~(\ref{eq:lorentzian_integral}). }
\label{fig:contour}
\end{figure}

In the lowest order approximation we find rates which are proportional to the
BCS-density of states leading to divergences at the gap edges. We can can
renormalize the rates by introducing a finite lifetime $(\gamma/\hbar)^{-1}$
in the exponents of Eq.~(\ref{eq:gamma+_n->n+1}) and
Eq.~(\ref{eq:gamma+_n->n-1}). Since we are neglecting coherences the imaginary
parts of the rates do not contribute to the dynamics of the system. For
example consider  the  integral appearing in Eq.~(\ref{eq:gamma+_n->n+1}):
\begin{equation}\label{eq:integral_renormalization1}
\begin{split}&
\text{Re}\bigg(\int_{-\infty}^{\infty}dE~\int_0^{\infty} dt_2 e^{\ih(E + \omega
+ i
\gamma)t_2}
f^{+}(E)\,D(E) \bigg) \\&
=   \int_{-\infty}^{\infty}dE~
\frac{ \hbar\gamma}{(E+\omega)^2+\gamma^2}\,
f^+(E)\, D(E),
\end{split}
\end{equation}
where we introduced $\omega = E_{n'm'} + \mu_{\eta}$.
Generalizing the integral for the cases ($N\rightarrow N \pm 1$) it reads
\begin{equation}\label{eq:lorentzian_integral}
\begin{split}
\hbar  \int_{-\infty}^{\infty}dE~
L(E,\omega)\,f^\pm(E)\, D(E) = \hbar  \int_{-\infty}^{\infty}dE\, F(E),
\end{split}
\end{equation}
where 
\begin{equation}
 L(E,\omega) = \frac{ \gamma}{(E+\omega)^2+\gamma^2}
\end{equation} 
describes the Lorentzian and $F(E)= L(E,\omega) \,f^\pm(E)\, D(E)$.
We can solve the integral of Eq.~(\ref{eq:lorentzian_integral}) using residue
calculus hence. To this extend we  analyze the  singularities of the
integrand and
the area in which the integrand is analytic. The Lorentzian $L(E,\omega)$  has
poles at
\begin{equation}
 E = -\omega \mp i \gamma,
\end{equation} 
with the corresponding residues:
\begin{equation}
 Res_{E=-\omega \mp \gamma}\, L(E) = \frac{\pm i }{2}.
\end{equation}
The poles of the Fermi function $ f^\pm(E)$ are  purely imaginary and equally
distributed along the imaginary axis:
\begin{equation}
 E= \frac{i \pi }{\beta} (2n + 1) \quad n \in \mathbb{Z},
\end{equation}
with the residues
\begin{equation}
 Res_{E= \frac{i \pi }{\beta} (2n + 1) }\, f^{\pm}(E) = \frac{\mp 1}{\beta}.
\end{equation}
The square roots in the BCS-density of states $D(E)$ have branch cuts
along the real axis.
In Fig.~\ref{fig:contour} we sketched the contour in the complex plane which is
slightly shifted away from the real axis with $\epsilon =1/R$. In the limit
$R\rightarrow \infty$ the integral along the semicircle vanishes and we are
left with:
\begin{equation}\label{eq:lorentzian_integral_2}
\begin{split}&
\lim_{R\rightarrow \infty} \int_{-R}^{R}dx F(x + i \epsilon) =  2\pi i
\sum_{\alpha}Res_{z=\alpha} F(z).
\end{split}
\end{equation} 
In the limit $R\rightarrow \infty$ Eq.~(\ref{eq:lorentzian_integral_2}) is
mapped back into the real integral of Eq.~(\ref{eq:lorentzian_integral}), and we
find:
\begin{equation}
 \begin{split}&
  \hbar  \int_{-\infty}^{\infty}dE~ L(E)\,f^\pm(E)\, D(E) \\ =& 
 \pi \hbar \text{Re}\bigg(
f^+(-\omega + i\gamma) \, D(E-\omega + i\gamma) \bigg).
 \end{split}
\end{equation}

\end{appendix}

\bibliographystyle{apsrev}  
\bibliography{bibliography}

\begin{thebibliography}{41}
\bibitem[{not()}]{note_groundstate}
\bibinfo{note}{The number of Cooper pairs is not fixed by the chemical
  potential since $\hat H_G$ is independent of it. Charging effects would
  physically determine the total number of electrons and correspondingly the
  number of Cooper pairs in $\ket{\text{GS}}$.}
\end{thebibliography}

\end{document}